\documentclass[english,reprint,aps,superscriptaddress,prl,hidelinks]{revtex4-1}
\usepackage{xr-hyper}
\usepackage{hyperref}
\usepackage{xr}

\let\oldref\ref
\makeatletter
\newcommand*{\addFileDependency}[1]{
  \typeout{(#1)}
  \@addtofilelist{#1}
  \IfFileExists{#1}{}{\typeout{No file #1.}}
}
\makeatother
\newcommand*{\myexternaldocument}[2][]{%
    \externaldocument[#1]{#2}%
    \addFileDependency{#2.tex}%
    \addFileDependency{#2.aux}%
}

\hypersetup{
colorlinks,
linktocpage=true,
    linkcolor={red!50!black},
    citecolor={blue!50!black},
    urlcolor={blue!80!black}
}

\newif\ifarxiv
\arxivtrue

\ifarxiv 
\newcommand{\newstuff}[1]{{#1}} 
\global\long\def\smref#1{\ref{app:#1}}

\global\long\def\smmref#1{~\ref{app:#1} in the Supplemental Material (SM)}

\else 
\global\long\def\smref#1{\ref{sm-app:#1}}

\global\long\def\smmref#1{~\ref{sm-app:#1} in the Supplemental Material (SM)~\cite{sm}}
\myexternaldocument[sm-]{sm}
\newcommand{\newstuff}[1]{{\color{Blue}{#1}}}
\fi

\makeatletter 
\global\long\def\mysection#1{\vspace{2pt}\emph{#1}.--- \@ifnextchar\par{\@gobble}{}}%
\makeatother

\input{preamble}
\usepackage[capitalize]{cleveref}
\usepackage{babel}
\begin{document}
\title{Information geometry of excess and housekeeping entropy production}
\author{Artemy Kolchinsky}
\affiliation{Universal Biology Institute, The University of Tokyo, 7-3-1 Hongo,
Bunkyo-ku, Tokyo 113-0033, Japan}
\author{Andreas Dechant}
\affiliation{Department of Physics No. 1, Graduate School of Science, Kyoto University,
Kyoto 606-8502, Japan}
\author{Kohei Yoshimura}
\affiliation{Department of Physics, The University of Tokyo, 7-3-1 Hongo, Bunkyo-ku,
Tokyo 113-0033, Japan}
\author{Sosuke Ito}
\affiliation{Universal Biology Institute, The University of Tokyo, 7-3-1 Hongo,
Bunkyo-ku, Tokyo 113-0033, Japan}
\affiliation{Department of Physics, The University of Tokyo, 7-3-1 Hongo, Bunkyo-ku,
Tokyo 113-0033, Japan}
\begin{abstract}
A nonequilibrium system is characterized by a set of thermodynamic
forces and fluxes which give rise to entropy production (EP). We show
that these forces and fluxes have an information-geometric structure,
which allows us to decompose EP into contributions from different
types of forces in general (linear and nonlinear) discrete systems.
We focus on the excess and housekeeping decomposition, which separates
contributions from conservative and nonconservative forces. Unlike
the Hatano-Sasa decomposition, our housekeeping/excess terms are always well-defined,
including in systems with odd variables and nonlinear systems without
steady states. Our decomposition leads to far-from-equilibrium thermodynamic
uncertainty relations and speed limits. \newstuff{As an illustration,
we derive a thermodynamic bound on the time necessary for one cycle
in a chemical oscillator.}
\end{abstract}
\maketitle

A major goal of nonequilibrium thermodynamics is to understand entropy
production (EP) from an operational point of view, in terms of tradeoffs
between EP and functional properties such as speed of dynamical evolution~\citep{aurell2011optimal, shiraishi_speed_2018}
and statistics of fluctuating observables~\citep{gingrich2016dissipation}.
However, EP can arise from different factors, including relaxation
from nonequilibrium states, nonconservative forces, and exchange of
conserved quantities between different reservoirs. In this Letter,
we use methods from information geometry~\citep{amari2016information,ay2017information}
to decompose EP into nonnegative contributions from different sources
and to study their operational consequences. 

We focus on the decomposition of EP into \emph{excess} and \emph{housekeeping}
terms~\citep{esposito2010three,hatano2001steady,oono1998steady}.
At a general level, excess EP is the contribution from conservative
forces, which arise from the change of a thermodynamic potential, and
it is expected to vanish in steady state. Housekeeping EP is the contribution
from nonconservative forces, such as the forces that generate cyclic fluxes in nonequilibrium
steady states. The housekeeping contribution can be arbitrarily large, and in
general it diverges during quasistatic transformations between nonequilibrium
steady states~\citep{mandal2016analysis,maes2014nonequilibrium}.
One of the main goals of this decomposition is to derive %
tighter thermodynamic tradeoffs and bounds by considering only the
excess part of EP~\citep{oono1998steady,maes2014nonequilibrium}.

While the housekeeping/excess decomposition is well understood at
a conceptual level, identifying the correct formal definitions remains
an open area of research~\citep{komatsu2008steady,spinney2012nonequilibrium,lee2013fluctuation,maes2014nonequilibrium,kohei2022,sagawa2011geometrical}.
The best known proposal is the Hatano-Sasa (HS) decomposition, also
called the adiabatic/nonadiabatic decomposition \citep{hatano2001steady,esposito2010three,esposito2007entropy,rao2016nonequilibrium,ge2016nonequilibrium}.
However, the HS decomposition has several drawbacks. First, its physical
meaning in terms of experimentally accessible observables is unclear~\citep{dechant2022geometric,dechant2022geometricCoupling,maes2014nonequilibrium}.
Second, it can lead to unphysical negative values
in stochastic systems with odd variables (variables such as velocity whose sign changes under time-reversal)~\citep{ford2012entropy,spinney2012nonequilibrium,lee2013fluctuation}.
It can also lead to negative values in chemical systems
that violate complex balance~\citep{rao2016nonequilibrium,ge2016nonequilibrium}.
Finally, it is unclear how to define the HS decomposition for systems
that lack stable steady states, such as chemical systems that exhibit
oscillations~\citep{kohei2022}. These drawbacks suggest that the
HS decomposition is not the ultimate definition of excess and housekeeping
EP.

Here we propose a new excess/housekeeping decomposition which resolves
all of these issues. %
Our decomposition is derived using techniques from information geometry,
and it is well-defined and nonnegative for all discrete systems, including
systems with odd variables and nonlinear chemical systems without
steady states. Our excess EP is experimentally accessible via statistics
of fluctuating observables, and it leads to new thermodynamic uncertainty
relations (TURs) and thermodynamic speed limits {(TSLs)}, which
can be tight even in the far-from-equilibrium regime.

Our approach is related to the decomposition proposed by Maes and
Netočný (MN) for Langevin systems~\citep{maes2014nonequilibrium}, which is recovered in the appropriate continuum limit. The MN decomposition was studied from a geometric perspective in Refs.~\citep{nakazato2021geometrical, dechant2022geometric,dechant2022geometricCoupling},
and generalized to discrete systems by the present authors in Ref.~\citep{kohei2022}.
Unlike these previous papers, which used a generalized Euclidean geometry,
here we consider the non-Euclidean setting of information geometry,
which is more appropriate for far-from-equilibrium systems (for a
comparison with Ref.~\citep{kohei2022}, see \smmref{ons}.).

\newstuff{This work complements existing research on geometry and
thermodynamics \citep{weinhold_metric_1975,ruppeiner_thermodynamics_1979,salamon_thermodynamic_1983,schlogl_thermodynamic_1985,janyszek_riemannian_1990,diosi_thermodynamic_1996},}
including recent studies of stochastic thermodynamics and information
geometry~\cite{mrugala_statistical_1990,brody_geometrical_1995,crooks_measuring_2007,sivak_thermodynamic_2012,ito2018stochastic,nakamura2019reconsideration,ito2020stochastic,ito2022information,kolchinsky2021work,yoshimura2021information,ohga2021information1, sughiyama2021hessian,ohga2021information,kobayashi2021kinetic,nicholson2018nonequilibrium,shiraishi2019information,van2021geometrical}.
However, almost all of these studies considered the geometry of thermodynamic
states, rather than dynamical quantities (thermodynamic forces and
fluxes) as pursued here. Exceptions include Ref.~\citep{ito2020unified},
which studied information geometry of trajectories in stochastic
systems, but did not derive decompositions based on constraints on
forces nor analyze their operational implications. Finally, Ref.~\citep{Kobayashi2022}
recently considered decompositions of fluxes and forces using different
type of information geometry (see \smref{numericalcomp} for a summary
and comparison).

\mysection{Setup}

We consider a system with $\numstate$ species or states with distribution
$\pp=(p_{1},\dots,p_{\numstate})\in\mathbb{R}_{>0}^{\numstate}$ at
time $t$. The system evolves in continuous time, either as a linear
stochastic master equation or a nonlinear rate equation (deterministic
chemical reaction system). The dynamics are generated by a set of
$\numedge$ reversible reactions, where each reaction $\rr\in\{1..\numedge\}$
is associated with a unique reverse reaction $\negedge\in\{1..\numedge\}$. The reactions are
also associated
with a set of (one-way) fluxes $\jj=(\TJ_{1},\dots,\TJ_{\numedge})\in\mathbb{R}_{>0}^{\numedge}$.
Note that $\pp$ and $\jj$ generally depend on time, though
we leave this dependence implicit in the notation. We make no assumptions
about the form of the fluxes (e.g., no assumption of mass action kinetics),
except where otherwise noted.

\newstuff{For each reaction $\rho$ and species $x$, the stoichiometric
coefficient $\BBgrad_{\rr x}=-\BBgrad_{\negedge x}\in\mathbb{Z}$ indicates
how many units of $x$ are added or removed by $\rr$. The overall
matrix $\BBgrad\in\mathbb{Z}^{\numedge\times\numstate}$ acts the
discrete gradient operator: for any state observable $\pot\in\mathbb{R}^{\numstate}$,
$[\BBgrad\pot]_{\rr}$ indicates how much reaction $\rr$ changes
the amount of $\pot$. Its transpose $\BBdiv$ acts as the (negative) discrete divergence operator.} The system's distribution evolves according
to the continuity equation $\dt\pp=\BBdiv\jj$, while expectations of observables evolve as 
$(\dt\pp)^{T}\pot=\jj^{T}\BBgrad\pot$.

The reactions are also associated with a set of 
thermodynamic forces $\ff=(f_{1},\dots,f_{\numedge})\in\mathbb{R}^{\numedge}$ which we assume obey local detailed balance.
 \newstuff{
 For now, we restrict
our attention to systems without odd variables, in which case the forces are given by the
log ratio of forward and reverse fluxes, $\ffrr=\ln (\jjrr/\jjEdgeNegRev)$. Note that $f_\rho$ is the change in total entropy due to reaction $\rho$ and the entropy production rate (EPR) is given by $\epr=\sum_\rho J_\rho f_\rho=\sum_{\rr}\jjrr\ln(\jjrr/\jjEdgeNegRev)$.
}

To make things concrete, consider a stochastic master equation without
odd variables. Here $\pp$ is a probability distribution which evolves
as $\dt p_{x}=\sum_{y(\ne x),\bath}(p_{y}R_{xy}^{\bath}-p_{x}R_{yx}^{\bath})$,
where $R_{yx}^{\bath}$ is the rate of transitions $x\shortrightarrow y$
mediated by reservoir $\bath$. Each ``reaction'' $\rr$ represents
one transition $(x\shortrightarrow y,\bath)$ with
flux $\jjrr=p_{x}R_{yx}^{\bath}$, stoichiometry $\BBgrad_{\rr z}=\delta_{zy}-\delta_{zx}$ (so that $[\nabla\pot]_{\rho}=\phi_{y}-\phi_{x}$), and 
reverse reaction $\negedge$ corresponding to the transition $(y\shortrightarrow x,\bath)$.


Alternatively, for deterministic chemical systems, $\pp$ is
a vector of nonnegative concentrations of different chemical species,
$\BBgrad$ is the transpose of the stoichiometric matrix, and $\jj$ is the vector of fluxes across reversible
reactions (see \smref{graph} for more details of chemical systems,
including a generalization of our formalism to account for external
currents).

To introduce techniques from information geometry, we define an exponential 
family of fluxes parameterized by $\ee\in\mathbb{R}^{\numedge}$:
\begin{align}
\jjrrT{\ee}:=\jjrr e^{\eerr-\ffrr}.\label{eq:pfam}
\end{align}
Within this family, the actual fluxes are recovered at $\ee=\ff$, $\jjrr=\jjrrT{\ff}$, and the reverse fluxes are recovered at $\ee=\zz$, $\jjEdgeNegRev=\jjrrT{\zz}$. 
The generalized Kullback-Leibler
(KL) divergence \citep[Def. 2.8, ][]{ay2017information}
provides an information-theoretic
distance between members of this family:
\begin{equation}
\DD(\ee\Vert\ee')=\sum_{\rr}\Big[\jjrrT{\ee}\ln\frac{\jjrrT{\ee}}{\jjrrT{\ee'}}+\jjrrT{\ee'}-\jjrrT{\ee}\Big]\ge0.\label{eq:klgen}
\end{equation}
Importantly, the EPR 
can be written using this KL divergence, 
\begin{align}
\epr=\DD(\ff\Vert\zz),\label{eq:eprdist}
\end{align}
which follows from Eqs.~(\ref{eq:pfam})-(\ref{eq:klgen}) and $\sum_{\rr}\jjrr=\sum_{\rr}\jjEdgeNegRev$.

\begin{figure}[b]
\includegraphics[width=0.9\columnwidth]{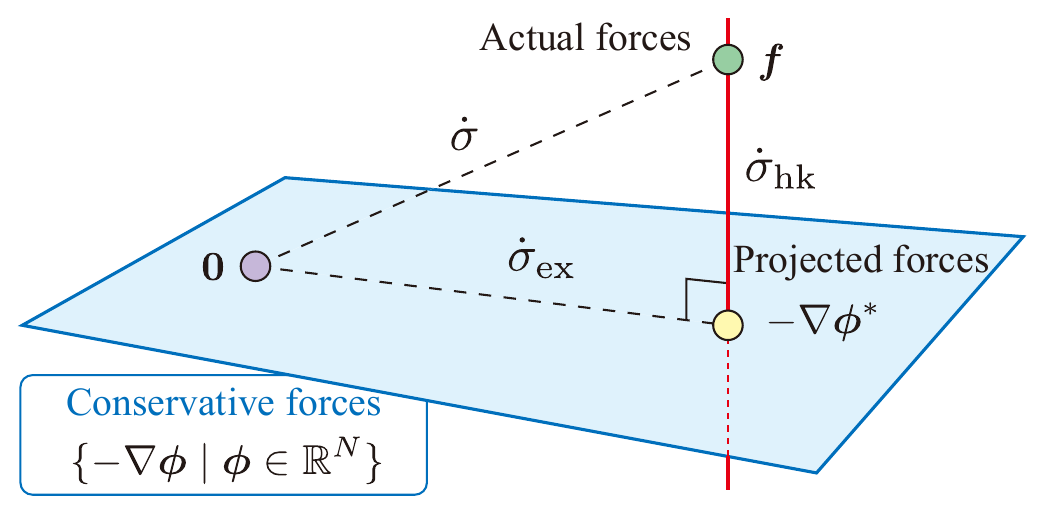}

\caption{\label{fig:pyth}Illustration of excess/housekeeping decomposition,
\cref{eq:pyth}. 
The red line indicates the set of parameter values that lead to the same dynamical evolution as the forward fluxes, $\jjT{\ee}=\nabla \jjT{\ff}=\nabla\jj$.
}
\end{figure}

\mysection{Housekeeping vs. excess EPR}

We now introduce our decomposition of the EPR, which is shown visually in \cref{fig:pyth}. Derivations of these
results, which use standard techniques from information geometry,
are in \smref{info-geom}.

Recall that a vector of thermodynamic forces $\ff$ is called \emph{conservative}
if it is the negative gradient of some thermodynamic potential $\pot\in\mathbb{R}^{\numstate}$,
so that $\ff=\negBBgrad\pot$. For example, for a master equation
that obeys detailed balance relative to an equilibrium distribution
$\bm{\pi}$ ($R_{xy}\pi_{y}=R_{yx}\pi_{x}$), the thermodynamic forces
are conservative for the potential $\phi_{x}=\ln(p_{x}/\pi_{x})$.

In general, the housekeeping EPR should vanish when $\ff$ is conservative.
Motivated by this, we define the housekeeping EPR as the information-theoretic
distance between $\ff$ and the closest conservative force $\negBBgrad\pot$,
\begin{align}
\eprHK:=\min_{\pot\in\mathbb{R}^{\numstate}}\DD(\ff\Vert\negBBgrad\pot)\ge0.\label{eq:hkdef}
\end{align}
Note that $0\le\eprHK\le\epr$, since $\DD$ is nonnegative and $\DD(\ff\Vert\zz)=\epr$
is achieved by $\pot=\zz$. \newstuff{Furthermore, the minimum is
always achieved by some optimal potential $\pot^{*}$, and the optimal
conservative force $\negBBgrad\pot^{*}$ is unique (\smref{info-geom-unq}).
Therefore, when $\ff$ is conservative, $\eprHK$ vanishes and $\negBBgrad\pot^{*}=\ff$.}

The excess EPR is defined as the remainder $\eprEX:=\epr-\eprHK$.
Using the duality principle from information geometry, $\eprEX$
can be written in a variational form (\smref{info-geom-dual}),
\begin{align}
\eprEX=\min_{\ee\in\mathbb{R}^{\numedge}}\DD(\ee\Vert\zz)\quad\text{where}\quad\BBdiv\jjT{\ee}=\dtp.\label{eq:maxent}
\end{align}
This means that $\eprEX$ is the distance from the closest $\jjT{\ee}$ to the
reverse fluxes $\jjT{\zz}$ such that $\jjT{\ee}$ leads to
the same dynamical evolution as the actual fluxes, $\dtp=\BBdiv \jj$. 
The optimum is achieved by the optimal conservative force $\negBBgrad\pot^{*}$ in \cref{eq:hkdef}. 
Both \cref{eq:hkdef,eq:maxent} are convex optimization
problems that can be solved using standard numerical techniques.


Combining these results, we can write our decomposition using the
\emph{Pythagorean relation for KL divergence} (see \cref{fig:pyth}):
\begin{align}
\underbrace{\DD(\ff\Vert\zz)}_{\begin{matrix}\epr\end{matrix}}=\underbrace{\DD(\ff\Vert\negBBgrad\pot^{*})}_{\begin{matrix}\eprHK\end{matrix}}+\underbrace{\DD(\negBBgrad\pot^{*}\Vert\zz)}_{\begin{matrix}\eprEX\end{matrix}}.\label{eq:pyth}
\end{align}
\cref{eq:pyth} is analogous to the Pythagorean Theorem in Euclidean
geometry, with the KL divergence playing the role of squared
Euclidean distance. 

The optimal potential $\pot^{*}$ has several interesting properties. 
Given \cref{eq:maxent}, the fluxes corresponding to the optimal conservative force $\negBBgrad\pot^{*}$
give rise to the actual dynamical evolution, 
$\dtp=\BBdiv \jjT{\negBBgrad\pot^{*}}$. Moreover, as we show in \smref{gradient}, this dynamical evolution can be written as a gradient flow for a free energy defined
in terms of $\pot^{*}$, which generalizes an existing result for
conservative forces~\citep{maas_gradient_2011,mielke2011gradient}. 
In essence, $\pot^{*}$ acts as the system's ``effective'' free energy,
which is well-defined even in the presence of nonconservative forces.

Although the definition of $\eprEX$ makes no explicit mention of
steady state, $\eprEX$ vanishes if the system is in steady state (\smref{info-geom-vanish}). Note that 
the reverse fluxes induce the opposite dynamics as the forward fluxes,
$\BBgrad\jjT{\zz}=-\BBgrad\jj$. In steady state, $\dtp=\BBgrad\jj=\zz$,
so the reverse fluxes satisfy the constraint in \cref{eq:maxent},
$\BBdiv\jjT{\zz}=\zz$, while achieving the minimum $\eprEX=0$.
 In addition, by properties of KL divergence,
$\eprEX\sim\Vert \dtp\Vert^{2}$ near steady state. 
Thus, the time integral of excess EP vanishes in the quasistatic limit
of slow driving, when $\tau\to\infty$ and $\dtp\sim1/\tau$. 

We can compare our decomposition to the HS decomposition. For stochastic
master equations, the HS housekeeping EPR can be expressed as
the information-theoretic distance between $\ff$ and a particular vector of conservative forces,
$\eprhsHK=\DD(\ff\Vert\negBBgrad\pot^{\mathrm{ss}})$, where the potential $\phi_{x}^{\mathrm{ss}}=\ln p_{x}/\pi_{x}^{\mathrm{ss}}$
is defined via the steady-state distribution $\bm{\pi}^{\mathrm{ss}}$
(\smref{hs}). The same result also holds for nonlinear chemical
systems with mass action kinetics and complex balance. The variational
principle in \cref{eq:hkdef} then implies that $\eprHK\le\eprhsHK$
and $\eprEX\ge\eprhsEX$. The remainder $\epr_{\text{cpl}}=\eprhsHK-\eprHK\ge0$
is a ``coupling term'', similar to one recently proposed
for Langevin dynamics~\citep{dechant2022geometricCoupling}.

Importantly, our general approach can be used to derive many other
kinds of decomposition of the EPR, not just the housekeeping/excess
decomposition. By replacing $\BBgrad$ with some other matrix in \cref{eq:hkdef},
one can consider projections onto a different subspace of forces,
rather than the set of conservative forces. We leave exploration
of such alternative decompositions for future work.

\newstuff{

\mysection{Excess EPR and dynamical fluctuations}

Our housekeeping/excess decomposition is directly related to the dynamical
fluctuations of observables, which provides an effective way to bound
and estimate $\eprEX$ from experimental data. This differs from the
existing decompositions, including the HS decomposition, which
requires knowledge of the steady state and has no direct relationship
with physical observables at a given point in time \citep{dechant2022geometric}.

Let us first review the relationship between dynamical fluctuations
and the EPR. It has been recently shown that the EPR obeys the following
variational principle \citep{kim2020learning,otsubo2020estimating}
(\smref{info-geom-legendre}):
\begin{align}
\epr=\max_{\ee\in\Theta}\sum_{\rr}J_{\rr}(\eerr-e^{-\eerr}+1),\label{eq:varEPR}
\end{align}
where $\Theta=\{\ee\in\mathbb{R}^{\numedge}:\eerr=\eerrR\,\forall\rr\}$
is the set of antisymmetric current observables, and the maximum is
achieved by the thermodynamic forces $\ff$. A series expansion gives
$\epr=\max_{\ee\in\Theta}[2\langle\ee\rangle-\sum_{k>1}(-1)^{k}\langle\ee^{k}\rangle/k!]$,
where $\langle\ee^{k}\rangle=\jj^{T}\ee^{k}$ is the $k$th moment
of $\ee$. Thus, in stochastic systems, EPR constrains the mean and
higher-order fluctuations of all current observables. This constraint
is stronger than standard TURs \citep{van2020entropy}, because the
maximum in \cref{eq:varEPR} always gives the exact EPR, including
in discrete systems and systems arbitrarily far from equilibrium and
steady state. This provides a powerful method for measuring
EPR from empirical observations \citep{kim2020learning,otsubo2020estimating},
since any choice of current observable gives a bound on the EPR which
can be made arbitrarily tight by optimizing over observables.

Our decomposition has a closely related interpretation. Specifically,
excess EPR can be written in terms of the following variational formula,
\begin{align}
\eprEX=\max_{\pot\in\mathbb{R}^{\numstate}}\sum_{\rr}J_{\rr}([\negBBgrad\pot]_{\rr}-e^{[\BBgrad\pot]_{\rr}}+1),\label{eq:legendre}
\end{align}
where the $\pot$ that achieves the maximum is the optimal
potential $\pot^{*}$ from \cref{eq:hkdef}.  \cref{eq:legendre} follows
from $\eprEX=\DD(\ff\Vert\zz)-\min_{\pot}\DD(\ff\Vert\negBBgrad\pot)$
and rearranging (\smref{info-geom-legendre}). 

This shows that $\eprEX$ satisfies the same variational principle as the EPR,
except that current observables are restricted to those of the form $\negBBgrad\pot$,
as generated by the change of some state observable $\pot$. (We use the symbol $\pot$
for both state potentials and state observables, since they are not
formally distinguished in our approach.) Thus,
in stochastic systems, $\eprEX$ constrains the dynamic fluctuations
of all state observables; conversely, $\eprEX$ is the part of EPR
that can be accessed by measuring the fluctuations of state observables.
The same techniques proposed in \citep{kim2020learning,otsubo2020estimating}
to estimate EPR can also be used to estimate excess EPR from real-world
data. In fact, it may be much easier to estimate $\eprEX$ than $\epr$,
since $\eprEX$ does not require measurements of arbitrary current
observables but only changes of state observables (i.e., by measuring some $\pot$ at time $t$
and $t+\delta t$ over many runs of a process).

For a system governed by conservative forces, $\epr=\eprEX$ and the
two variational principles in \cref{eq:varEPR,eq:legendre} agree.
In fact, for stochastic master equations with conservative forces,
Ref.~\citep{shiraishi2019information} derived a variational expression
of $\epr$ that turns out to be equivalent to \cref{eq:legendre}.
 Our decomposition generalizes that variational principle to linear
and nonlinear systems with nonconservative forces. It also
generalizes the main result of Ref.~\citep{shiraishi2019information},
which is an information-theoretic bound on the speed of evolution
in stochastic equations with time-symmetric 
driving (\smref{variational}).}

\mysection{TURs and TSLs}

We now use \cref{eq:legendre} to derive TURs and TSLs for the excess
EP. Our results apply both to linear and nonlinear systems.

Consider any state observable $\obs$, and assume without loss of
generality that it is scaled so that $\left\Vert \BBgrad\obs(t)\right\Vert _{\infty}\le1$.
Our bounds are stated in terms of the observable's speed $\chngObs:=\sum_{\rr}\jjrr[\BBgrad\obs]_{\rr}=(\dtp)^{T}\obs$
and the dynamical activity $A=\sum_{\rr}\jjrr$ \newstuff{(the overall
number of reactions per second \citep{shiraishi_speed_2018})}. We
also consider the ``mean deviation'' of the observable's dynamic
fluctuations, $\actOBS=\sum_{\rr}\jjrr\vert[\BBgrad\obs]_{\rr}\vert\le A$.
$\actOBS$ is a nonnegative measure of the size of fluctuations that
vanishes when $\obs$ is a conserved quantity.

We first derive the following short-time TUR,
\begin{align}
\eprEX\ge2\chngObs\,\atanh\frac{\chngObs}{\actOBS}\ge2\chngObs\,\atanh\frac{\chngObs}{A}.\label{eq:tur1}
\end{align}
This result follows from \cref{eq:legendre} and a simple bound on the
exponential function, with all details in \smref{tur}. We can also derive a finite-time version of \cref{eq:tur1}. We consider
a process over time $t\in[0,\tau]$ and any time-dependent observable
$\pot(t)$ ($\left\Vert \BBgrad\obs(t)\right\Vert _{\infty}\le1$
at all $t$). Using \cref{eq:tur1} and Jensen's inequality, we derive a bound on the integrated excess EP $\epEX(\tau)=\int_0^\tau \eprEX(t) \,dt$,
\begin{align}
\epEX(\tau)\ge2\mathcal{L}_{\pot}\,\atanh\frac{\mathcal{L}_{\pot}}{\tau\langle\actOBS\rangle}\ge2\mathcal{L}_{\pot}\,\atanh\frac{\mathcal{L}_{\pot}}{\tau\langle A\rangle},\label{eq:inttur2}
\end{align}
where $\mathcal{L}_{\pot}=\int_{0}^{\tau}\vert\chngObs(t)\vert\,dt$
is the trajectory length of the observable, while $\langle\actOBS\rangle=(1/\tau)\int_{0}^{\tau}\actOBS(t)\,dt$
and $\langle A\rangle=(1/\tau)\int_{0}^{\tau}\sum_{\rr}\jjrr(t)\,dt$
are time-averaged mean deviation and dynamical activity. A simple
rearrangement of \cref{eq:inttur2} gives a far-from-equilibrium TSL:
\begin{align}
\tau\ge\frac{\LLL_{\pot}}{\langle\actOBS\rangle}\coth\frac{\Sigma_{\mathrm{ex}}}{2\LLL_{\pot}}\ge\frac{\LLL_{\pot}}{\langle A\rangle}\coth\frac{\Sigma_{\mathrm{ex}}}{2\LLL_{\pot}}.\label{eq:timelimit}
\end{align}
Naturally, these bounds also hold for total EP, $\epEX\le\ep$.

The choice of the time-dependent observable in \cref{eq:inttur2,eq:timelimit}
can be used to derive various specialized TSLs. For example, for 
the ``total variation'' observable $\phi_{x}^{\mathrm{tv}}(t):=(1/2)\mathrm{sign}(\dt p_{x}(t))$, the trajectory length is 
$\mathcal{L}_{\pot^{\mathrm{tv}}}=(1/2)\int_{0}^{\tau}\left\Vert \dt\pp\right\Vert _{1}\,dt$. \newstuff{Alternatively, using a different time-dependent observable, one can derive
TSLs for the ``$L_{1}$-Wasserstein path length'', an important
quantity in optimal transport theory (see \citep[Sec.  4,][]{dechant2022minimum}
for details).}

\begin{figure}[b]
\includegraphics[width=1\columnwidth]{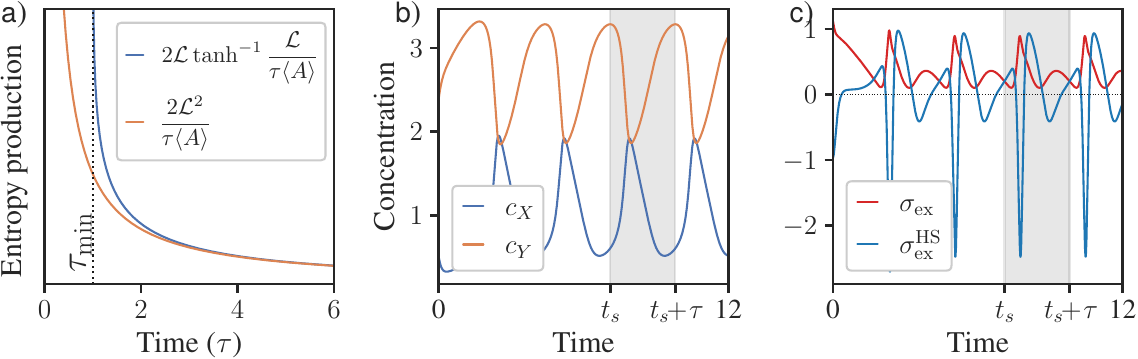}

\caption{\label{fig:bruss}a) Finite-time scaling in \cref{eq:inttur2} vs.
$1/\tau$ scaling in standard TSLs. \newstuff{b) Trajectory of
a Brusselator model that reaches a limit cycle. We derive a bound
on the minimal time $\tau$ need to complete a cycle. c) $\protect\eprEX(t)$
vs. $\protect\eprhsEX(t)$ for the Brusselator model.}}
\end{figure}

The bound in \cref{eq:tur1} diverges as $\chngObs/A\to1$, the absolutely
irreversible regime where all activity is channeled into directed
movement. It is stronger than conventional
TURs that are quadratic in $\chngObs$ \citep{horowitz2020thermodynamic},
which are tight only near equilibrium and do not diverge in the
limit of absolute irreversibility. Furthermore, \cref{eq:timelimit}
implies a finite minimal time $\taumin=\LLL/\langle A\rangle$, where
$\epEX$ diverges as $-\ln(\tau-\taumin)$ as $\tau\to\tau_{\min}$.
This is stronger than the $1/\tau$ finite-time scaling reported in
conventional TSLs~\citep{aurell2011optimal,shiraishi_speed_2018,van2021geometrical,nakazato2021geometrical,yoshimura2021thermodynamic,hamazaki2022speed,kohei2022},
which only become tight in the limit of slow driving~\citep{berut2012experimental,zhen2021universal,zhen2022inverse}.
The difference between these finite-time scaling relations is shown
in \cref{fig:bruss}(a). Our bounds can be related to recently
proposed far-from-equilibrium TURs~\citep{delvenne2021thermo,van2020unifiedCSLTUR}
and TSLs~\citep{dechant2022minimum,lee2022speed,salazar2022lower} but go beyond these existing results, 
which either do not apply to nonlinear chemical systems and/or do not
separately consider excess EP (therefore cannot be tight in
the presence of nonconservative forces).


\newstuff{

\mysection{Example}

We illustrate our results on the Brusselator~\cite{prigogine1968symmetry}, a well-known model of
an autocatalytic chemical system. The model contains three reactions:
1) $\varnothing\rightleftarrows X$, 2) $X\rightleftarrows Y$, and
3) $2X+Y\rightleftarrows3X$. We assume mass action kinetics with
rate constants $k_{1}^{+}=k_{1}^{-}=k_{2}^{-}=k_{3}^{-}=1$ and $k_{2}^{+}=15,k_{3}^{+}=5$. 
For these parameters, the system exhibits limit cycle behavior.

The time-dependent concentrations $c_{X}(t),c_{Y}(t)$ are shown in \cref{fig:bruss}(b), with
one cycle period $t\in[t_{s},t_{s}+\tau]$ highlighted. \cref{fig:bruss}(c) shows $\eprEX(t)$ at different times; it is always nonnegative and tends to be large when the concentrations are changing rapidly.
\cref{fig:bruss}(c) also shows the HS excess EPR $\eprhsEX(t)$ calculated using the (unstable) fixed point $(c_{X}^{*},c_{Y}^{*})=(1,8/3)$. It can be seen that $\eprhsEX$ sometimes exhibits unphysical negative
values.

Next, we illustrate \cref{eq:timelimit} by deriving a TSL on the cycle
period $\tau$, stated in terms of the cycle arc length $\mathcal{C}=\int_{t_{s}}^{t_{s}+\tau}\left\Vert \dot{\cc}(t)\right\Vert _{2}dt$,
 dynamical activity $\langle A\rangle=(1/\tau)\int_{t_{s}}^{t_{s}+\tau}\sum_{\rr}J_{\rr}(t)\,dt$
(average number of reactions/second), and excess EP, $\epEX(\tau)=\int_{t_{s}}^{t_{s}+\tau}\eprEX(t)\,dt$.
We bound the cycle length using the total variation observable $\pot^{\mathrm{tv}}$, $\mathcal{L}_{\pot^{\mathrm{tv}}}=(1/2)\int_{t_{s}}^{t_{s}+\tau}\left\Vert \dt {\cc}(t)\right\Vert _{1}dt\ge \mathcal{C}/2$,
where we used the inequality between $\ell_{1}$ and $\ell_{2}$ norms.
Since \cref{eq:timelimit} is monotonically decreasing in $\mathcal{L}_{\pot}$, 
we recover the following TSL:
\begin{align}
\tau\ge\tau_{\min}^{\mathrm{(1)}}=(\mathcal{C}/2\langle A\rangle)\coth(\Sigma_{\mathrm{ex}}/\mathcal{C}).\label{eq:cycleTSL}
\end{align}
Remarkably, this bound is not specific to the Brusselator, and actually
applies to any limit cycle in any chemical system.

We compare our TSL to a weaker bound that uses the EP rather than excess EP, $\tau_{\min}^{(2)}=(\mathcal{C}/2\langle A\rangle)\coth(\Sigma/\mathcal{C})$
(which is also new to the literature). Finally, we compare our result
to an existing TSL for chemical systems, proposed in
Ref.~\citep{yoshimura2021thermodynamic}. Using Eq.~12 and Eq.~13 in
that paper, along with $\int_{t_{s}}^{t_{s}+\tau}\left\Vert \dot{\cc}(t)\right\Vert _{1}dt\ge\mathcal{C}$,
implies the bound $\tau\ge\mathcal{C}^{2}/\langle\tilde{D}\rangle\Sigma$,
where the quantity $\langle\tilde{D}\rangle$ depends on system fluxes
and stoichiometry~\citep{yoshimura2021thermodynamic}. For the Brusselator,
$\langle \tilde{D}\rangle \le\langle A\rangle$,  which finally gives $\tau_{\min}^{(3)}=\mathcal{C}^{2}/\langle A\rangle\Sigma$.

Using numerically calculated values of $\tau$, $\mathcal{C}$, $\langle A\rangle$,
and $\epEX$ and $\ep$, we compare the tightness of the bounds:
\[
\tau/\tau_{\min}^{\mathrm{(1)}}\approx0.13\qquad\tau/\tau_{\min}^{\mathrm{(2)}}\approx0.026\qquad\tau/\tau_{\min}^{\mathrm{(3)}}\approx0.0013.
\]
Thus, despite its generality, \cref{eq:cycleTSL} provides a relatively
tight bound on the cycle period, which is two orders of magnitude
better than any previously known bound.

}

\newstuff{

\mysection{Odd variables}

We finish by discussing our decomposition in the context of linear
master equations with odd variables. In such systems, the thermodynamic force across 
 a single transition $x\to y$ is $f_{yx} = \ln [p_{x}R_{yx}/(p_{y}R_{\oddconj x\oddconj y})]$, where $\pp$ is the
probability distribution, $R$ is the rate matrix, and $\oddconj x$
is state $x$ with odd variables flipped in sign~\citep{lee2013fluctuation,spinney2012entropy}. 
Odd variables lead to problems with the HS decomposition, such as
negative values of housekeeping EPR~\citep{spinney2012nonequilibrium,ford2012entropy,lee2013fluctuation}. To our knowledge, no universally applicable housekeeping/excess decomposition has been proposed for such systems. 

On the other hand, our decomposition generalizes immediately to systems
with odd variables, as long as the vector of thermodynamic forces is defined appropriately.
In the presence of odd variables, the EPR does not have the usual form $\epr = \sum_{\rr}\jjrr \ffrr$. Nonetheless, as we show in \smref{odd}, it can still be written as 
the generalized
KL divergence $\epr=\DD(\ff\Vert\zz)$, as in \cref{eq:eprdist}. Our expressions of $\eprEX$ and
$\eprHK$ in \cref{eq:hkdef} and \cref{eq:maxent}, as well as the
Pythagorean relation in \cref{eq:pyth}, hold without modification.
In \smref{odd-example}, we consider an example system with odd variables
and demonstrate that our decomposition gives physically meaningful
values, even when the HS housekeeping EPR is negative.

Note that some of our other results must be qualified in the presence
of odd variables. For instance, 
excess EPR vanishes in steady state only if the steady-state
distribution obeys time-reversal symmetry $\pi_{x}^{\mathrm{ss}}=\pi_{\oddconj x}^{\mathrm{ss}}$,
and the same holds for the bound $\eprEX\le\eprhsEX$. Finally, the
variational principle in \cref{eq:legendre}, as well as the TUR and
TSL derived from it, do not hold in general for systems with odd variables.

}

%
%
%
%
%

\vspace{5pt}

\ifarxiv  
    \let\savedaddcontentsline\addcontentsline
    \renewcommand\addcontentsline[3]{}
\fi  

\begin{acknowledgments}
S. I. thanks Masafumi Oizumi for fruitful discussions. A.~D.~is
supported by JSPS KAKENHI Grants No.~19H05795, and No.~22K13974.
K.~Y.~is supported by Grant-in-Aid for JSPS Fellows (Grant No.~22J21619).
S.~I.~is supported by JSPS KAKENHI Grants No.~19H05796, No.~21H01560,
and No.~22H01141, and UTEC-UTokyo FSI Research Grant Program. 
\end{acknowledgments}


\bibliographystyle{IEEEtran}
\bibliography{writeup}

\ifarxiv
  \let\addcontentsline\savedaddcontentsline
  \clearpage
  \newcommand{\mainref}[1]{Eq.~(\ref*{#1})}
  
\onecolumngrid

\renewcommand\thesection{SM\arabic{section}}
\renewcommand\thesubsection{\arabic{subsection}}
\renewcommand\thesubsubsection{\alph{subsubsection}}
\renewcommand\theequation{S\arabic{equation}}
\makeatletter 
\renewcommand*{\p@subsection}{\thesection.} 
\makeatother 

\titleformat*{\section}{\fontsize{14}{20}\sffamily\bfseries\selectfont} 
\titleformat{\subsection}[hang]{\fontsize{12}{17}\sffamily\bfseries\selectfont}{\thesection.\thesubsection}{10pt}{}{}
\titleformat{\subsubsection}[hang]{\fontsize{11}{15}\sffamily\itshape\selectfont}{\thesection.\thesubsection}{10pt}{}{}

\begin{center}

{\fontsize{16pt}{22bp}\bfseries Information geometry of excess and housekeeping entropy production }

\vspace{20pt}

{\fontsize{14pt}{15} Artemy Kolchinsky, Andreas Dechant, Kohei Yoshimura, and Sosuke Ito }

\vspace{20pt}
{\fontsize{15pt}{22bp}\bfseries Supplementary Material }

\end{center}

\fontsize{11pt}{18bp}\selectfont

\tableofcontents{}

\fontsize{11pt}{22bp}\selectfont 
\clearpage

\global\long\def\TJ{J}%
\global\long\def\jyx{\TJ_{yx}}%
\global\long\def\jxy{\TJ_{xy}}%
\global\long\def\jxyRev{\tilde{\TJ}_{xy}}%
\global\long\def\eq{\mathrm{eq}}%
\global\long\def\sss{\mathrm{ss}}%
\global\long\def\peq{\bm{\pi}^\eq}%
\global\long\def\pss{\bm{\pi}^{\sss}}%

\global\long\def\dtpx{\dt p_{\xx}}%
\global\long\def\eprHKons{\epr_{\text{hk}}^{\text{ons}}}%
\global\long\def\eprEXons{\epr_{\text{ex}}^{\text{ons}}}%
\global\long\def\netflux{\bm{\mathcal{J}}}%
\global\long\def\netjR{\mathcal{J}_{r}}%

\global\long\def\xx{x}%
\global\long\def\yy{y}%
\global\long\def\xy{xy}%
\global\long\def\yx{yx}%
\global\long\def\BBxxrr{\BBgrad_{\rr\xx}}%
\global\long\def\BBxxmrr{\BBgrad_{\negedge\xx}}%
\global\long\def\barBBdiv{\bar{\BBgrad}^{T}}%
\global\long\def\barBBgrad{\bar{\BBgrad}}%
\global\long\def\gstd{\Delta G_{r}^{\circ}}%
\global\long\def\jjrrCRN{\TJ_{r}^{\rightarrow}}%
\global\long\def\jjRevrrCRN{\TJ_{r}^{\leftarrow}}%


\section{Deterministic chemical systems}

\label{app:graph} 
Here we show how our formalism can be used to analyze deterministic
chemical systems, and how the continuity equation $\dtp=\BBdiv\jj$
gives the deterministic rate equation.

Consider a chemical system with $\numstate$ species and $m$ reversible
reactions. Let the $r\in\{1..m\}$ reversible reaction be 
\begin{align}
\sum_{\xx}\nu_{\xx r}Z_{\xx}\rightleftarrows\sum_{\xx}\kappa_{\xx r}Z_{\xx},\label{ap-chemeq1}
\end{align}
where $Z_{\xx}$ is the $x$-th species, and $\nu_{\xx r}$ and $\kappa_{\xx r}$
are stoichiometric coefficients. We also write the forward and reverse
flux across this reaction as $\jjrrCRN$ and $\jjRevrrCRN$. As an example,
for a chemical system with mass action kinetics, these fluxes are
given by 
\begin{align}
\jjrrCRN=k_{r}^{\rightarrow}\prod_{\xx}c_{\xx}^{\nu_{\xx r}},\qquad\jjRevrrCRN=k_{r}^{\leftarrow}\prod_{\xx}c_{\xx}^{\kappa_{\xx r}}.\label{eq:massaction}
\end{align}
where $k_{r}^{\rightarrow}$ and $k_{r}^{\leftarrow}$ are the forward
and reverse rate constants and $c_{\xx}$ is the concentration of
$Z_{\xx}$. (Note that mass action kinetics are used as an example;
our results do not assume mass action kinetics except where explicitly
stated.)

To connect to the formalism described in the main text, each reversible
reaction should be treated as two separate one-way reactions $\rho$ and $\negedge$,
with fluxes and stoichiometric entries defined as:
\begin{align}
\jjrr=\jjrrCRN,\quad\BBxxrr=\nu_{\xx r}-\kappa_{\xx r}\qquad\qquad
\jjEdgeNegRev=\jjRevrrCRN,\quad \BBxxmrr=\kappa_{\xx r}-\nu_{\xx r} .
\end{align}
Thus, $m$ reversible reactions in the original representation give rise to $\numedge=2m$ one-way reactions in our formalism. 
Using these definitions, the deterministic rate equation is 
\begin{align}
\dt c_{\xx} & =\sum_{r=1}^{m}(\nu_{\xx r}-\kappa_{\xx r})(\jjrrCRN-\jjRevrrCRN)=\sum_{\rr=1}^{\numedge}\BBxxrr\jjrr.\label{ap-rateeq}
\end{align}
This recovers the continuity equation mentioned in the main text,
$\dtp=\BBdiv\jj$, if we adopt the notation $p_{\xx}=c_{\xx}$.

Observe that our definition of the EPR coincides with the conventional
one for chemical reaction networks: 
\begin{align}
\dot{\sigma} & =\sum_{r=1}^{m}\jjrrCRN\ln\frac{\jjrrCRN}{\jjRevrrCRN}+\sum_{r=1}^{m}\jjRevrrCRN\ln\frac{\jjRevrrCRN}{\jjrrCRN}=\sum_{r=1}^{m}(\jjrrCRN-\jjRevrrCRN)\ln\frac{\jjrrCRN}{\jjRevrrCRN}.
\end{align}

We remark that a slightly different convention is used in Ref.~\citep{kohei2022}.
There, the notation $\mathbb{S}$ is used instead of $\BBdiv$, and
each reversible reaction $r$ is treated as a single ``edge'' $r$
with net flux $\netjR=\jjrrCRN-\jjRevrrCRN$ (which may be positive
or negative). In that paper, $m$ reversible reactions lead to $M=m$
edges, with the associated rate equation $\dt c_{\xx}=\sum_{r=1}^{m}\mathbb{S}_{\xx r}\netjR$.

We finish by showing that our approach immediately generalizes to
chemical systems subject to external currents, such as a continuous-flow
stirred-tank reactor with inflow and dilution. In this case, the dynamical
evolution obeys the modified continuity equation
\[
\dtp=\BBdiv\jj+\cc,
\]
where $\cc\in\mathbb{R}^{N}$ is a vector of external currents (like
$\pp$ and $\jj$, in general $\cc$ can depend on time). In this
case, the expression of the EPR in terms the generalized KL divergence
remains unmodified, $\epr=\DD(\ff\Vert\zz)$ as in \mainref{eq:eprdist}
in the main text, as does the definition of the housekeeping EPR in
\mainref{eq:hkdef} and the Pythagorean decomposition in \mainref{eq:pyth}.
The expression of the excess EPR in \mainref{eq:maxent}
should be written in a slightly more general way, 
\begin{align}
\eprEX & =\inf_{\ee\in\mathbb{R}^{\numedge}}\DD(\ee\Vert\zz)\quad\text{where}\quad\BBdiv\jjT{\ee}=\BBdiv\jj\label{eq:maxent-2},
\end{align}
that is using $\BBdiv\jj$ instead of $\dtp$. We note that some of
the subsequent results, such as the statement that $\eprEX$ vanishes
in steady state, do not necessarily hold in the presence of external
currents.

\section{Information-geometric fundamentals}

\label{app:info-geom}

Our decomposition of the entropy production uses existing techniques
from information geometry (see in particular Theorem 1 in \citep{collins2002logistic}).
To be self-contained, in this appendix we provide simple derivations
of our main results.


\subsection{Existence and uniqueness of minimizer in the definition of the housekeeping
EPR, \mainref{eq:hkdef}}

\label{app:info-geom-unq}

We first demonstrate the existence of the minimizer of the optimization
problem which defines housekeeping EPR, \mainref{eq:hkdef} in the
main text.

To begin, write the generalized KL divergence from \mainref{eq:klgen}
as 
\begin{align}
\DD(\ff\Vert\ee)=\sum_{\rr}\jjrr(\ffrr-{\eerr}+e^{{\eerr}-\ffrr}-1).\label{eq:appgenkl}
\end{align}
Note that the function $\eerr\mapsto\ffrr-{\eerr}+e^{{\eerr}-\ffrr}-1$
is nonnegative, continuous, strictly convex, and ``coercive'' (diverges
to $\infty$ as $\vert\eerr\vert\to\infty$). By assumption $\jjrr>0$
for all $\rr$, therefore the function $\ee\mapsto\DD(\ff\Vert\ee)$
is also nonnegative, continuous, strictly convex, and coercive. Continuity
and coercivity imply that the sublevel set $A:=\{\ee\in\mathbb{R}^{\numedge}:\DD(\ff\Vert\ee)\le\DD(\ff\Vert\zz)\}$
is compact~\citep[Lemma~8.3]{calafiore2014optimization}. The set
$B:=A\cap\mathrm{im}\,\negBBgrad$ is nonempty (it contains $\zz$)
and also compact, since it is the intersection of a compact set and
a closed set \citep[p. 38,][]{rudinPrinciplesMathematicalAnalysis1976}.
Finally, by the extreme value theorem, there exists some $\ff^{*}=\negBBgrad\pot^{*}\in B$
such that 
\[
\DD(\ff\Vert\negBBgrad\pot^{*})=\inf_{\ee\in B}\DD(\ff\Vert\ee)=\inf_{\ee\in\mathrm{im}\,\negBBgrad}\DD(\ff\Vert\ee)=\inf_{\pot\in\mathbb{R}^{\numstate}}\DD(\ff\Vert\negBBgrad\pot).
\]
This proves that the minimum is attained.

Note that the optimal potential $\pot^{*}$ may not be unique, because
$\BBgrad\pot^{*}=\BBgrad(\pot^{*}+\mathbf{v})$ for any null vector
$\mathbf{v}$ of $\BBgrad$ (physically, such null vectors represent
conserved quantities). Nonetheless, the optimal conservative force
$\ff^{*}=\negBBgrad\pot^{*}$ is always unique due to strict convexity
of the function $\ee\to\DD(\ff\Vert\ee)$. 

\subsection{Dual variational principle for excess EPR, \mainref{eq:maxent}}

\label{app:info-geom-dual}

Here we derive the dual variational principle for the excess EPR,
which appears as the maximization problem in \mainref{eq:maxent}
in the main text. 

Consider the partial derivatives of the objective in \mainref{eq:hkdef}
in the main text,
\begin{align*}
\partial_{\phi_{\xx}}\DD(\ff\Vert\negBBgrad\pot)
&=\sum_{\rr}\jjrr\big(\BBxxrr- e^{[\negBBgrad \pot]_\rr -\ffrr}\BBxxrr\big) = [\BBdiv\jj-\BBdiv\jjT{\negBBgrad\pot}]_{\xx},
\end{align*}
where we used \cref{eq:appgenkl} and $\partial_{\phi_{\xx}}[\negBBgrad \pot]_\rr=-\BBxxrr$.  
The partial derivatives vanish for all $\xx$ at the minimizer $\pot^{*}$,
so 
\begin{align}
\BBdiv\jj=\BBdiv\jjT{\negBBgrad\pot^{*}}.\label{eq:appg4}
\end{align}

\noindent Next, write the optimization problem in \mainref{eq:maxent}
using the equivalent space of strictly positive fluxes, 
\begin{align}
\eprEX= & \inf_{\jj'\in\mathbb{R}_{>0}^{\numedge}}D(\jj'\Vert\jjT{\zz})\quad\text{where}\;\;\BBdiv\jj'=\dtp,\label{eq:maxent-fluxes}
\end{align}
where $D$ is the generalized KL divergence for flux vectors, 
\[
D(\jj'\Vert\jjT{\zz}):=\sum_{\rr}\jjrr'\ln\frac{\jjrr'}{\jjrrT{\zz}}-\jjrr'+\jjrrT{\zz}.
\]
For convenience, let $\ff^{*}=\negBBgrad\pot^{*}$ indicate the optimal
conservative force in \mainref{eq:hkdef}. Since $\dtp=\BBdiv\jj$ and $\BBdiv\jj=\BBdiv\jjT{\ff^{*}}$
from \cref{eq:appg4}, $\jjT{\ff^{*}}$ is in the feasible set of \cref{eq:maxent-fluxes}.
Now consider any other $\jj'$ that satisfies $\BBdiv\jj'=\dtp$,
and define the convex mixture $\bm{a}(\lambda):=(1-\lambda)\jjT{\ff^{*}}+\lambda\jj'$.
The directional derivative of the objective function in \cref{eq:maxent-fluxes}
at $\jjT{\ff^{*}}$ toward $\jj'$ is given by 
\begin{align*}
\frac{\partial}{\partial\lambda}D(\bm{a}(\lambda)\Vert\jjT{\zz})\Big\vert_{\lambda=0} & =\sum_{\rr}(\jjrr'-\jjrrT{\ff^{*}})\ln\frac{a_\rr(\lambda)\vert_{\lambda=0}}{\jjrrT{\zz}}\\
& =\sum_{\rr}(\jjrr'-\jjrrT{\ff^{*}})\ln\frac{\jjrr e^{\ffrr^*-\ffrr}}{\jjrr e^{-\ffrr}}\\
 & =\sum_{\rr}(\jjrr'-\jjrrT{\ff^{*}})f_{\rr}^{*}\\
 & ={\ff^{*}}^{T}(\jj'-\jjT{\ff^{*}})=-{\pot^{*}}^{T}\BBdiv(\jj'-\jjT{\ff^{*}}).
\end{align*}
This directional derivative vanishes since $\BBdiv\jj'=\dtp=\BBdiv\jjT{\ff^{*}}$.
Because this holds for every $\jj'$ and the KL divergence is convex in both arguments~\citep{ay2017information}, $\jjT{\ff^{*}}$ is the solution of the
optimization problem in \mainref{eq:maxent}.

\subsection{Pythagorean relation for housekeeping and excess EPR, \mainref{eq:pyth}}

\label{app:info-geom-pyth}

Here we derive the Pythagorean relation for housekeeping and excess
EPR, \mainref{eq:pyth} in the main text.

Recall that the excess EPR is defined as 
\begin{align}
\eprEX & :=\DD(\ff\Vert\zz)-\DD(\ff\Vert\negBBgrad\pot^{*}).\label{eq:appz2}
\end{align}
Using \cref{eq:appgenkl}, the definition of $\jjT{\ee}$, and a
bit of rearranging, this can be rewritten as 
\begin{align}
\eprEX & =\sum_{\rr}\jjrr\big([\negBBgrad\pot^{*}]_{\rr}+ e^{-\ffrr}- e^{[\negBBgrad\pot^{*}]_{\rr}-\ffrr}\big).\label{eq:appz3}
\end{align}
Next, observe that \cref{eq:appg4} implies that 
\begin{align*}
\sum_{\rr}\jjrr[\negBBgrad\pot^{*}]_{\rr}=\sum_{\rr}\jjrr e^{[\negBBgrad\pot^{*}]_{\rr}-\ffrr}[-\BBgrad\pot^{*}]_{\rr}.
\end{align*}
Plugging back into \cref{eq:appz3} gives 
\begin{align}
\eprEX & =\sum_{\rr}\jjrr e^{[\negBBgrad\pot^{*}]_{\rr}-\ffrr}\big([\negBBgrad\pot^{*}]_{\rr}+e^{[\BBgrad\pot^{*}]_{\rr}}-1\big)=\DD(\negBBgrad\pot^{*}\Vert\zz).\label{eq:appz4}
\end{align}
The Pythagorean relation follows from \cref{eq:appz2,eq:appz4}.

\subsection{Excess EPR as a maximization problem, \mainref{eq:legendre}}

\label{app:info-geom-legendre}

Here we derive \mainref{eq:legendre} in the main text, which represents
excess EPR as a maximization problem. 

First, we use our definition of the excess and housekeeping EPR to
write 
\begin{align}
\eprEX=\epr-\eprHK=\DD(\ff\Vert\zz)-\min_{\pot\in\mathbb{R}^{\numstate}}\DD(\ff\Vert\negBBgrad\pot)=\max_{\pot\in\mathbb{R}^{\numstate}}\,\big(\DD(\ff\Vert\zz)-\DD(\ff\Vert\negBBgrad\pot)\big).\label{eq:app-l0}
\end{align}
Using \cref{eq:appgenkl} and the definition of $\jjT{\ee}$, the
objective on the RHS can be written as 
\begin{align*}
\DD(\ff\Vert\zz)-\DD(\ff\Vert\negBBgrad\pot) & =\sum_{\rr}\jjrr(\ffrr+e^{-\ffrr}-1)-\sum_{\rr}\jjrr(\ffrr-[\negBBgrad\pot]_{\rr}+e^{[\negBBgrad\pot]_{\rr}-\ffrr}-1).
\end{align*}
Cancelling terms and plugging back into \cref{eq:app-l0} gives
\begin{align}
\eprEX & =\max_{\pot\in\mathbb{R}^{\numstate}}\sum_{\rr}\jjrr[\negBBgrad\pot]_{\rr}-\sum_{\rr}\jjrr e^{-\ffrr}(e^{[\negBBgrad\pot]_{\rr}}-1).\label{eq:app-l1}
\end{align}
The variational expression in \cref{eq:app-l1} is completely general
and does not make use of any physical assumptions. 

To derive \mainref{eq:legendre},
we now introduce the assumption that there are no odd variables. As
described in the main text, this means that forces are defined as $\ffrr = \ln (\jjrr/\jjEdgeNegRev)$, where $\negedge$ is the reverse reaction corresponding
to reaction $\rr$, thus the fluxes obey
$\jjrr e^{-\ffrr}=\jjEdgeNegRev$. Note also that in all systems, the stoichiometry
of the reverse reactions obeys $\BBxxmrr=-\BBxxrr$, thus $[\negBBgrad\pot]_{\rr}=[\BBgrad\pot]_{\negedge}$.
We now perform a change of variables $\rr\to\negedge$
in the last sum in \cref{eq:app-l1} and rearrange. This gives \mainref{eq:legendre}:
\begin{align}
\eprEX=\max_{\pot\in\mathbb{R}^{\numstate}}\sum_{\rr}\jjrr([\negBBgrad\pot]_{\rr}-e^{[\BBgrad\pot]_{\rr}}+1).\label{eq:applegendre}
\end{align}

We emphasize that a similar technique can be used to derive the variational
expression of the overall EPR, $\epr$, found in \mainref{eq:varEPR} in the main text.
Specifically, let us write 
\begin{align}
\epr=\DD(\ff\Vert\zz)-\min_{\ee\in\Theta}\DD(\ff\Vert\ee)=\max_{\ee\in\Theta}\,[\DD(\ff\Vert\zz)-\DD(\ff\Vert\ee)],\label{eq:app-l0-1}
\end{align}
where $\Theta \subset \mathbb{R}^\numedge$ is the set of antisymmetric current observables.  This expression holds because $\DD(\ff\Vert\ee)=0$ when $\ee=\ff\in\Theta$. 
By expanding and rearranging terms, in the same manner as above, we arrive
at 
\begin{align}
\epr & =\max_{\ee\in\Theta}\sum_{\rr}\jjrr\eerr-\sum_{\rr}\jjrr e^{-\ffrr}(e^{\eerr}-1).\label{eq:app-l1-1}
\end{align}
This is a variational expression that holds without any assumptions.
In the information theory literature, it is sometimes called the Donsker-Varadhan
representation of the KL divergence. 

As above, however, we may now introduce the assumption
that there are no odd variables, so that $\jjrr e^{-\ffrr} = \jjEdgeNegRev$. Using this assumption, along with
the fact that $\ee$ is antisymmetric, we perform a change of variables
$\rr\to \negedge$ in the last sum in \cref{eq:app-l1-1} and rearrange.
This gives \mainref{eq:varEPR} in the main text,
\begin{align}
\epr=\max_{\ee\in\Theta}\sum_{\rr}J_{\rr}(\eerr-e^{-\eerr}+1).\label{eq:dfd3}
\end{align}
We note that \cref{eq:dfd3} has previously appeared in Ref.~\citep{otsubo2022estimating}
in the context of stochastic systems with linear dynamics. However,
as our derivation shows, the same variational expression also applies
to nonlinear chemical systems.

\subsection{Excess EPR vanishes in steady state and scales as $\Vert\mathrm{d}_{t}\bm{p}\Vert^2$ near steady state}

\label{app:info-geom-vanish}

Here we show that for systems without odd degrees of freedom, $\eprEX$ vanishes in steady state. More generally, we show that $\eprEX\sim\Vert\dtp\Vert^{2}$ near steady state.

To show that $\eprEX$ vanishes in steady state, we use the relation
\begin{align}
\BBdiv \jjT{\zz}=-\BBdiv \jj,\label{eq:app-reverse}
\end{align} which follows from 
$$
[\BBdiv \jjT{\zz}]_x = \sum_\rr \BBxxrr\jjrrT{\zz}=\sum_\rr \BBxxrr\jjrr e^{-\ffrr}
=\sum_\rr \BBxxrr\jjEdgeNegRev=\sum_\rr \BBxxmrr\jjrr=-\sum_\rr \BBxxrr\jjrr=-[\BBdiv \jj]_x.
$$
Here we used the definition of $\jjrrT{\zz}$ from \mainref{eq:pfam} (main text), $\ffrr = \ln (\jjrr/\jjEdgeNegRev)$ (local detailed balance for systems without odd variables), changed variables as $\rr\to \negedge$, and then applied the stoichiometric identity $\BBxxrr=-\BBxxmrr$. Therefore, if a system is in steady state, $\BBdiv \jj = 0 = -\BBdiv \jjT{\zz}$, so $\jjT{\zz}$ satisfies the constraint in \mainref{eq:maxent} while achieving the minimum value $\DD(\zz\Vert \zz)=0$. 

To show that $\eprEX\sim\Vert\dtp\Vert^{2}$, we use the variational principle from \mainref{eq:maxent} in
the main text, via its equivalent formulation in terms of strictly
positive fluxes as \cref{eq:maxent-fluxes}. First, define the following
vector of fluxes,
\[
\jj'=:\jjT{\zz}+2{(\BBdiv)}^{+}(\dtp)=\jjT{\zz}+2{(\BBdiv)}^{+}\BBdiv\jj,
\]
where ${(\BBdiv)}^{+}$ is the pseudo-inverse of $\BBdiv$. Note that
\begin{align}
\Vert\jj'-\jjT{\zz}\Vert\le2\Vert{(\BBdiv)}^{+}\Vert\Vert\dtp\Vert,\label{eq:bnd3}
\end{align}
so for sufficiently small $\left\Vert \dtp\right\Vert $, it must
be that the element of $\jj'$ are strictly positive (since the elements
of $\jjT{\zz}$ are strictly positive). Next, observe that the fluxes $\jj'$ satisfy the constraint in \cref{eq:maxent-fluxes}:
\[
\BBdiv\jj'=\BBdiv\jjT{\zz}+2\BBdiv{(\BBdiv)}^{+}\BBdiv\jj=-\BBdiv\jj+2\BBdiv\jj=\BBdiv\jj=\dtp,
\]
where we used that $\BBdiv \jjT{\zz}=-\BBdiv \jj$. 
\cref{eq:maxent-fluxes} then implies that  
\begin{align}
0\le\eprEX\le D(\jj'\Vert\jjT{\zz}).\label{eq:ffff}
\end{align}
Finally, note that $D(\jj'\Vert \jjT{\zz})$ as a function of its first
argument is convex, differentiable, and achieves its minimum value
of 0 if $\jj'=\jjT{\zz}$ --- therefore it vanishes to first order in $\Vert\jj'-\jjT{\zz}\Vert$. \cref{eq:bnd3} implies that $\Vert\jj'-\jjT{\zz}\Vert$
is of order $\left\Vert \dtp\right\Vert $, so $D(\jj'\Vert \jjT{\zz})$
is of order $\Vert\dtp\Vert^{2}$. Then, \cref{eq:ffff} implies that $\eprEX$
is also of order $\Vert\dtp\Vert^{2}$.

\section{Generalization of variational principle from Ref.~\NoCaseChange{\citep{shiraishi2019information}}}

\label{app:variational}


Ref.~\citep{shiraishi2019information} showed that, for stochastic
master equations without odd variables and subject to conservative
forces, the EPR can be expressed in terms of a variational principle.
Here we demonstrate that our expression for the excess EPR, \cref{eq:app-l1},
provides a generalization of this variational principle to arbitrary
stochastic master equations, including ones with odd variables and
with nonconservative forces.

Consider a system whose probability distribution evolves according
to a stochastic master equation, 
\[
\dtpx(t)=\sum_{\yy(\ne x),\bath}(p_{\yy}(t)R_{\xy}^{\bath}-p_{\xx}(t)R_{\yx}^{\bath}),
\]
where $R_{\xy}^{\bath}$ is the rate of jumps $y\to x$ mediated by
reservoir $\bath$. Suppose that the system is also associated with
a set of reverse transition rates $R_{\oddconj\yy\oddconj\xx}^{\bath}$,
where $\oddconj$ indicates conjugation of odd variables (see Section~\oldref{app:odd}).

We now show that \cref{eq:app-l1} implies the following variational
principle for the excess EPR: 
\begin{align}
\eprEX=\max_{\qq\in\Omega}\,[-\dt\dKL(\pp(t)\Vert\qq(-t))].\label{eq:shi-gen}
\end{align}
where $\dKL$ is the KL divergence between normalized probability
distributions and $\Omega$ is the set of all probability distributions
over the states. The notation $\qq(-t)$ indicates that $\qq$ evolves
backwards in time under the reverse rates, 
\begin{align}
-\dt q_{\xx}(-t)=\sum_{\yy(\ne x),\bath}(q_{\yy}(-t)R_{\oddconj\xx\oddconj\yy}^{\bath}-q_{\xx}(-t)R_{\oddconj\yy\oddconj\xx}^{\bath}).\label{eq:gdd2}
\end{align}
\cref{eq:shi-gen} implies that $\eprEX$ is the fastest rate of contraction
of KL divergence between the actual distribution $\pp$ evolving forward in
time and any other distribution evolving backward in time under the
reverse rates. %
The maximum in \cref{eq:shi-gen} is achieved by the ``pseudo-equilibrium''
distribution $q_{\xx}^{*}\propto p_{\xx}e^{-\phi_{\xx}^{*}}$, defined
via the optimal potential $\pot^{*}$ in \mainref{eq:hkdef}.

To derive \cref{eq:shi-gen}, we defined a reaction $\rr$ for each
transition $(y\to x,\alpha)$, with backward and reverse fluxes $\jxy^{\bath}=p_{\yy}R_{\xy}^{\bath}$
and $\jxyRev^{\bath}=p_{x}R_{\oddconj\yy\oddconj\xx}^{\bath}$. (In
the special case of a system without odd variables, $\oddconj x=x$
and the backward fluxes involve only a time-reversal, $\jxyRev^{\bath}=p_{x}R_{\yy\xx}^{\bath}$).
We then apply \mainref{eq:legendre},
\begin{align*}
\eprEX & =\max_{\pot\in\mathbb{R}^{\numstate}}\Big[-\sum_{\xx}(\dtpx)\phi_{x}-\!\sum_{\xx\ne\yy,\bath}p_{y}R_{\oddconj x\oddconj y}^{\bath}(e^{\phi_{\xx}-\phi_{\yy}}-1)\Big].
\end{align*}
Next, we change the variable of optimization from potentials to probability
distributions $\qq\in\Omega$ via $\ln q_{\xx}=\ln p_{x}-\phi_{\xx}+\text{const}$.
Using this replacement, we rewrite the right hand side as 
\begin{align*}
 & \max_{\qq\in\Omega}\Big[-\sum_{\xx}(\dtpx(t))\ln\frac{p_{\xx}}{q_{\xx}}-\sum_{\xx\ne\yy,\bath}p_{y}R_{\oddconj x\oddconj y}^{\bath}\Big(\frac{p_{x}}{q_{\xx}}\frac{q_{y}}{p_{y}}-1\Big)\Big].
\end{align*}
Since $\sum_{x}\dtpx(t)=0$, the first sum is 
\begin{align}
 & -\sum_{\xx}(\dtpx(t))\ln\frac{p_{\xx}}{q_{\xx}}=-\sum_{\xx}(\dtpx(t)\ln p_{\xx})+\sum_{\xx}(\dtpx(t))\ln q_{\xx}.\label{eq:app-var-totalderiv}
\end{align}
We rewrite the second sum as 
\begin{align}
\sum_{\yy\ne x,\bath}\Big(\frac{p_{\xx}}{q_{\xx}}q_{\yy}R_{\oddconj x\oddconj y}^{\bath}-p_{y}R_{\oddconj x\oddconj y}^{\bath}\Big) & =\sum_{\yy\ne x,\bath}\Big(\frac{p_{\xx}}{q_{\xx}}q_{\yy}R_{\oddconj x\oddconj y}^{\bath}-p_{x}R_{\oddconj y\oddconj x}^{\bath}\Big)\nonumber \\
 & =\sum_{x}\frac{p_{\xx}}{q_{\xx}}\sum_{\yy(\ne x),\bath}\big(q_{\yy}R_{\oddconj x\oddconj y}^{\bath}-q_{x}R_{\oddconj y\oddconj x}^{\bath}\big)\nonumber \\
 & =-\sum_{x}\frac{p_{\xx}}{q_{\xx}}\dt q_{\xx}(-t)=-p_{\xx}\dt\ln q_{\xx}(-t),\label{eq:app-rev-evo}
\end{align}
\cref{eq:shi-gen} follows by combining \cref{eq:app-var-totalderiv,eq:app-rev-evo}
and rearranging.

For systems without odd variables and subject only to conservative
forces,
\[
\eprEX=\epr,\qquad R_{\oddconj y\oddconj x}^{\bath}=R_{yx}^{\bath},\qquad\qq^{*}=\peq,
\]
where $\peq$ is the stationary equilibrium distribution. In this
case, \cref{eq:shi-gen} reduces to Eq.~(2) in Ref.~\citep{shiraishi2019information}.

Our result can be used to derive the following bound,
\begin{align}
\epEX(\tau)\ge\dKL(\pp(0)\Vert\pp(\tau)),
\label{eq:app-info-relax-bound}
\end{align} 
which generalizes the main result of Ref.~\citep{shiraishi2019information}.
The derivations proceeds in the same way as in Ref.~\citep{shiraishi2019information},
Eq.~(3). Consider a system that undergoes a driving protocol $R(t)$
over $t\in[0,\tau]$ starting from some initial distribution $\pp(0)$,
giving rise to a trajectory of probability distributions $\{\pp(t):t\in[0,\tau]\}$.
Suppose that the system does not have odd variables and that the driving
protocol is time-symmetric, $R(t)=R(\tau-t)$. We can then choose
$\qq(0)=\pp(\tau)$ in \cref{eq:shi-gen}, so that $\qq(t)=\pp(\tau-t)$
is a solution to \cref{eq:gdd2}. We integrate both sides of \cref{eq:shi-gen}
from $t=0$ to $t=\tau/2$ to give
\[
\epEX(\tau/2)=\int_{0}^{\tau/2}\eprEX(t)\,dt\ge\int_{0}^{\tau/2}-\dt\dKL(\pp(t)\Vert\pp(\tau-t))\,dt=\dKL(\pp(0)\Vert\pp(\tau)).
\]
Since $\eprEX(t)\ge0$ at all $t$, we then have 
\[
\epEX(\tau)\ge\epEX(\tau/2)\ge\dKL(\pp(0)\Vert\pp(\tau)).
\]

\section{Optimal potential and gradient flow}

\label{app:gradient}
\global\long\def\grad{\mathrm{grad}}%

It is known that, for a system without odd variables and subject only
to conservative forces, the temporal evolution can be expressed as
the gradient flow of a free energy potential. This result has been
shown both for stochastic master equations~\citep{maas_gradient_2011,van2021geometrical}
and for chemical systems with mass action kinetics~\citep{mielke2011gradient}.
We briefly review these results in our own notation.

Consider some distribution potential $\Phi(\pp)$ defined over the
system's distribution $\pp$, which may be a normalized probability
distribution or an unnormalized concentration vector. The time derivative
of this function is given by
\begin{align}
\partial_{t}\Phi(\pp(t))=(\dtp(t))^{T}\grad_{\pp}\,\Phi(\pp(t)).\label{eq:gf1}
\end{align}
Note that  we typically leave the time dependence of $\pp(t)$ implicit, writing it as $\pp$.  Note also that we write the gradient as $\grad_{\pp}\,\Phi(\pp):=(\partial_{p_1} \Phi(\pp),\dots, \partial_{p_\numstate} \Phi(\pp))$, rather than $\nabla \Phi(\pp)$, to avoid confusion with the discrete gradient matrix 
used in other parts of this work.  A system is said to evolve according
to a gradient flow if 
\begin{align}
\dtp(t)=-K\grad_{\pp}\,\Phi(\pp)\label{eq:gf2}
\end{align}
for some positive-semidefinite matrix $K$. Note that $K$ can depend
on time, though we omit this in our notation. Combining \cref{eq:gf1,eq:gf2}
implies 
\begin{align}
\partial_{t}\Phi(\pp(t))=-(\grad_{\pp}\,\Phi(\pp))^{T}K(\grad_{\pp}\,\Phi(\pp))\le0,\label{eq:grad3}
\end{align}
thus the value of $\Phi(\pp(t))$ decreases over time.

Let us now suppose that the system has only conservative forces and
undergoes autonomous driving. We then define the distribution potential
as the generalized KL divergence between the system's actual distribution
and the equilibrium,
\[
\Phi(\pp):=D(\pp\Vert\peq)=\sum_{\xx}p_{x}\ln\frac{p_{x}}{\pi_{\xx}^{\eq}}-p_{x}+\pi_{\xx}^{\eq},
\]
which represents the free energy. It can then be shown that the temporal
evolution is a gradient flow \citep{maas_gradient_2011,mielke2011gradient},
\begin{align}
\dtp(t)=-K\grad_{\pp}\,D(\pp\Vert\peq),\label{eq:appgrad-1}
\end{align}
where $K$ is a $\numstate\times\numstate$ positive-semidefinite
matrix defined as $K=\BBdiv L\BBgrad$, where $L\in\mathbb{R}_{+}^{\numedge\times\numedge}$
is a diagonal matrix with entries $L_{\rr\rr}=\frac{1}{2}(\jjrr-\jjEdgeNegRev)/\ffrr$.
In fact, $L$ is a $\numedge\times\numedge$ Onsager matrix that maps forces to
net fluxes at the level of individual (one-way) reactions, while $K$ is a $\numstate\times\numstate$ Onsager matrix which maps forces to dynamics at the level of species. If the system is autonomous (no time-dependent
driving), then the equilibrium distribution $\peq$ and the function $D(\cdot\Vert\peq)$
do not depend on time. Therefore, free energy $D(\cdot\Vert\peq)$
is a Lyapunov function for the dynamics, which implies stability of
autonomous systems with conservative forces.

Our decomposition generalizes \cref{eq:appgrad-1} to  systems with nonconservative forces (and without odd variables). We show this using a similar
technique as found in Ref.~\citep{kohei2022}. First, define the
parameterized reaction-level Onsager matrix $\mathcal{L}(\ee)\in\mathbb{R}_{+}^{\numedge\times\numedge}$
as
\begin{align}
\mathcal{L}_{\rr\rr}(\ee)=\frac{1}{2}\jjrr e^{-\ffrr}(e^{\eerr}-1)/\eerr=\frac{1}{2} (\jjrrT{\ee}-\jjrrT{\zz})/\eerr.
\end{align}
This Onsager matrix maps forces to the net fluxes at the level of
individual edges, where the forward fluxes defined by the exponential
family in \mainref{eq:pfam} in the main text, 
\begin{align}
\mathcal{L}(\ee)\ee=\frac{1}{2}(\jjT{\ee}-\jjT{\zz}).\label{eq:d3-1}
\end{align}
Note that $\mathcal{L}(\ee)$ reduces to the previous Onsager matrix
when $\ee=\ff$, $L=\mathcal{L}(\ff)$. We also define a ``pseudo-equilibrium''
distribution using the optimal potential $\pot^{*}\in\mathbb{R}^{\numstate}$
from \mainref{eq:hkdef},
\begin{align}
\pi_{x}^{*}:=p_{x}e^{-\phi_{x}^{*}}.\label{eq:dddd}
\end{align}
Note that $\pot^{*}$ can always be chosen so that $\bm{\pi}^{*}$
satisfies the system's conservation laws (e.g., so that $\bm{\pi}^{*}$
is a normalized probability distribution in a stochastic system, satisfies
mass conservation in a chemical system, etc.), and in a system with
only conservative forces, $\pot^{*}$ can be chosen so that $\bm{\pi}^{*}=\peq$.
Then, in analogy to \cref{eq:appgrad-1}, the temporal evolution is
a gradient flow for the generalized KL divergence between $\pp$ and
$\bm{\pi}^{*}$,
\begin{align}
\dtp=\BBdiv\jj=\frac{1}{2}\BBdiv(\jjT{\negBBgrad\pot^{*}}-\jjT{\zz})=-\BBdiv\mathcal{L}(\negBBgrad\pot^{*})\BBgrad\pot^{*}=-\mathcal{K}\grad_{\pp}\,D(\pp\Vert\bm{\pi}^{*}),\label{eq:grnd}
\end{align}
where $\mathcal{K}:=\BBdiv\mathcal{L}(\negBBgrad\pot^{*})\BBgrad$
is a positive-semidefinite species-level Onsager matrix. In deriving
this result, we first used $\BBdiv\jjT{\negBBgrad\pot^{*}}=\BBdiv\jj$ from \cref{eq:appg4} 
and $\BBdiv\jjT{\zz}=-\BBdiv\jj$ from \cref{eq:app-reverse}, and then used \cref{eq:d3-1}.
 
We emphasize that this result holds for all systems, including ones
with nonconservative forces. Moreover, by \cref{eq:grad3}, this result
means that the distribution $\pp$ moves in time so as to decrease
$D(\pp\Vert\bm{\pi}^{*})$. However, because the pseudo-equilibrium
distribution $\bm{\pi}^{*}$ can itself depend on $\pp$, even in
an autonomous system, the function $\pp\to D(\cdot\Vert\bm{\pi}^{*})$
in general is not time-independent. This means that for systems with
nonconservative forces, \cref{eq:grnd} is in general a non-autonomous
gradient flow, and does not imply Lyapunov stability.

We finish by noting that a similar gradient flow result was also derived
in Ref.~\citep{kohei2022}. However, that result was based on a different
optimal potential (specifically, it was the optimal potential $\pot_{\text{ons}}^{*}$
from \cref{eq:proj1}, discussed in Section~\oldref{app:ons} above),
as well as a different Onsager matrix.

\section{Thermodynamic uncertainty relations}

\label{app:tur}

Here we provide a derivation of the thermodynamic uncertainty relations
(TURs), \mainref{eq:tur1} and \mainref{eq:inttur2}.

We first derive the TUR in \mainref{eq:tur1} in the main text. Let $\obs\in\mathbb{R}^{\numstate}$
be any state observable that satisfies the scaling condition $\left\Vert \BBgrad\obs\right\Vert _{\infty}=\max_{\rr}\vert[\BBgrad\obs]_{\rr}\vert\le1$.
We restrict \mainref{eq:legendre} in the main text to scalar multiples of $\obs$
and rearrange to give 
\begin{align}
\eprEX & \ge\max_{\lambda\in\mathbb{R}}\big[-\lambda\chngObs-\sum_{\rr}\jjrr(e^{\lambda[\BBgrad\pot]_{\rr}}-1)\big],\label{eq:legendre-tur}
\end{align}
where we used the definition $\chngObs=(\dtp)^{T}\obs=\jj^{T}\BBgrad\pot$.
Note that $[\BBgrad\pot]_{\rr}\in[-1,1]$ by the scaling assumption
and that $e^{\lambda x}-1\le x(e^{\lambda}-1)$ for $x\in[0,1]$ and
$e^{\lambda x}-1\le-x(e^{-\lambda}-1)$ for $x\in[-1,0]$. Plugging
these inequalities into \cref{eq:legendre-tur} leads to the bound
\begin{align}
\eprEX\ge\max_{\lambda\in\mathbb{R}}(-\lambda\chngObs+\actOBS-\actOBS^{+}e^{\lambda}-\actOBS^{-}e^{-\lambda}),\label{eq:leg3}
\end{align}
where we defined the positive ($\actOBS^{+}$) and negative ($\actOBS^{-}$)
activity of the observable as
\[
\actOBS^{+}:=\sum_{\rr:[\BBgrad\obs]_{\rr}>0}\jjrr[\BBgrad\obs]_{\rr}=(\actOBS+\chngObs)/2,\qquad\actOBS^{-}:=\sum_{\rr:[\BBgrad\obs]_{\rr}<0}\jjrr[-\BBgrad\obs]_{\rr}=(\actOBS-\chngObs)/2.
\]
\cref{eq:leg3} can be maximized in closed form to find the optimal
$\lambda^{*}=\ln\frac{\actOBS-\chngObs}{\actOBS+\chngObs}$. Plugging
this into \cref{eq:leg3} gives the first inequality in \mainref{eq:tur1}.
The second inequality follows by noting that $\actOBS\le A$.

We now derive the finite-time TUR in \mainref{eq:inttur2}. For notational
convenience, define $\Phi(x):=x\atanh x$. Then, write \mainref{eq:tur1}
as 
\[
\eprEX(t)\ge2\actOBS(t)\Phi\left(\frac{\chngObs(t)}{\actOBS(t)}\right)=2\actOBS(t)\Phi\left(\frac{\vert\chngObs(t)\vert}{\actOBS(t)}\right).
\]
We can then bound the integrated excess EP as 
\begin{align}
\epEX(\tau):=\int_{0}^{\tau}\eprEX(t)\,dt\ge2\int_{0}^{\tau}\actOBS(t)\Phi\left(\frac{\vert\chngObs(t)\vert}{\actOBS(t)}\right)\,dt=2\tau\langle\actOBS\rangle\int_{0}^{\tau}\frac{\actOBS(t)}{\tau\langle\actOBS\rangle}\Phi\left(\frac{\vert\chngObs(t)\vert}{\actOBS(t)}\right)\,dt.\label{eq:f2}
\end{align}
Applying Jensen's inequality to the convex function $\Phi$ gives
\[
\int_{0}^{\tau}\frac{\actOBS(t)}{\tau\langle\actOBS\rangle}\Phi\left(\frac{\vert\chngObs(t)\vert}{\actOBS(t)}\right)\,dt\ge\Phi\left(\frac{\int_{0}^{\tau}\vert\chngObs(t)\vert\,dt}{\tau\langle\actOBS\rangle}\right)\equiv\Phi\left(\frac{\LLL_{\pot}}{\tau\langle\actOBS\rangle}\right).
\]
Combining these results and the definition $\Phi$ gives 
\[
\epEX(\tau)\ge2\LLL_{\pot}\tanh^{-1}\frac{\LLL_{\pot}}{\tau\langle\actOBS\rangle},
\]
which is the first inequality in \mainref{eq:inttur2}. The second
inequality follows from $\langle\mathcal{A}\rangle\le\langle A\rangle$.

\section{Systems with odd variables}

\label{app:odd}

\subsection{Entropy production rate}

Here we consider stochastic jump process with odd variables, that
is variables such as velocity whose sign must be flipped under time
reversal. We derive an expression of EPR by starting from a discrete-time
formulation. 

Consider a system with odd variables that is coupled to a single heat
bath and evolves over some small time interval $\tau\ll1$. Let $p_{x}$
indicate the probability of state $\xx$ at time $t$, and let $T_{y\vert x}(\tau)$
indicate the conditional probability that the system is in state $y$
at time $t+\tau$, given that it was in state $x$ at time $t$.

For systems with odd variables, the entropy production (EP) of jump
from $\xx\to\yy$ is \citep[Eq.  11,][]{spinney2012entropy},
\begin{align}
\sigma_{\yx}(\tau)=\ln\frac{p_{\xx}T_{y\vert x}(\tau)}{p_{\yy}T_{\oddconj x\vert\oddconj y}(\tau)}.\label{eq:conjldb}
\end{align}
where $\oddconj\xx$ is the conjugation of state $\xx$ with odd-parity
variables flipped in sign. \cref{eq:conjldb} is the statement of the thermodynamic principle of local detailed balance for systems with odd variables (see \citep{spinney2012entropy,spinney2012nonequilibrium,lee2013fluctuation}
and \citep[Sec. 5.3.4, ][]{gardinerHandbookStochasticMethods2004}).
The expected EP over time $\tau$ is given by the KL divergence between
the forward and backward transition distributions,
\begin{align}
\ep(\tau)=D_{KL}(p_{x}T_{y\vert x}(\tau)\Vert p_{y}T_{\oddconj x\vert\oddconj y}(\tau))=\underbrace{\sum_{x\ne y}p_{\xx}T_{y\vert x}(\tau)\sigma_{\yx}(\tau)}_{\text{Transitions}}+\underbrace{\sum_{x}p_{\xx}T_{x\vert x}(\tau)\ln\frac{p_{\xx}T_{x\vert x}(\tau)}{p_{x}T_{\oddconj x\vert\oddconj x}(\tau)}}_{\text{Diagonals}}.\label{eq:derivapp}
\end{align}
The second term, which we label ``Diagonals'', vanishes if the system
doesn't have odd variables (since then $x=\oddconj x$).

The EPR is the derivative of EP with respect to $\tau$, $\epr=\partial_{\tau}\ep(\tau)$.
To compute this derivative, we suppose the transition matrix $T_{t,t+\tau}$
arises from a continuous-time Markov chain with a time-homogeneous
generator $R$, so
\begin{align}
T_{y\vert x}(\tau) & =\tau R_{\yx}+O(\tau^{2}),\qquad T_{x\vert x}(\tau)=1-\sum_{y(\ne x)}T_{y\vert x}(\tau)=1-\tau\sum_{\yy(\ne\xx)}R_{\yx}+O(\tau^{2}).\label{eq:df3-1}
\end{align}
We then evaluate the time derivative of EP at $\tau=0$, considering
the derivatives of the two terms in \cref{eq:derivapp} separately.
The first term gives
\begin{align}
\partial_{\tau}\sum_{x\ne y}p_{\xx}T_{y\vert x}(\tau)\sigma_{\yx}(\tau) & =\sum_{x\ne y}p_{x}R_{\yx}\ln\frac{p_{\xx}T_{y\vert x}(\tau)}{p_{\yy}T_{\oddconj x\vert\oddconj y}(\tau)}+p_{x}R_{\yx}-\frac{p_{\xx}T_{y\vert x}(\tau)}{p_{\yy}T_{\oddconj x\vert\oddconj y}(\tau)}p_{\yy}R_{\oddconj x\oddconj y}\nonumber \\
 & =\sum_{x\ne y}p_{x}R_{\yx}\ln\frac{p_{x}R_{\yx}}{p_{\yy}R_{\oddconj x\oddconj y}},\label{eq:htf4}
\end{align}
where we also used that $\lim_{\tau\to0}\frac{p_{\xx}T_{y\vert x}(\tau)}{p_{\yy}T_{\oddconj x\vert\oddconj y}(\tau)}=\frac{p_{x}R_{\yx}}{p_{\yy}R_{\oddconj x\oddconj y}}$.
The second term \cref{eq:derivapp} gives
\begin{align}
\partial_{\tau}\sum_{x}p_{\xx}T_{x\vert x}(\tau)\ln\frac{p_{\xx}T_{x\vert x}(\tau)}{p_{x}T_{\oddconj x\vert\oddconj x}(\tau)} & =\sum_{x}\partial_{\tau}p_{\xx}T_{x\vert x}(\tau)\ln\frac{p_{\xx}T_{x\vert x}(\tau)}{p_{x}T_{\oddconj x\vert\oddconj x}(\tau)}+\partial_{\tau}p_{\xx}T_{x\vert x}(\tau)-\frac{p_{x}T_{\oddconj x\vert\oddconj x}(\tau)}{p_{\xx}T_{x\vert x}(\tau)}\partial_{\tau}p_{x}T_{\oddconj x\vert\oddconj x}(\tau)\nonumber \\
 & =\sum_{x}\partial_{\tau}p_{\xx}T_{x\vert x}(\tau)-\partial_{\tau}p_{x}T_{\oddconj x\vert\oddconj x}(\tau)\nonumber \\
 & =-\sum_{x\ne y}p_{x}R_{\yx}+\sum_{x\ne y}p_{x}R_{\oddconj y\oddconj x}.\label{eq:sdf3}
\end{align}
In deriving this expression, we used that $\frac{p_{\xx}T_{x\vert x}(\tau)}{p_{x}T_{\oddconj x\vert\oddconj x}(\tau)}=1$
at $\tau=0$ as well as \cref{eq:df3-1}. Combining \cref{eq:htf4}
and \cref{eq:sdf3}, plus a bit of rearranging, gives
\begin{align}
\epr:=\partial_{\tau}\ep(\tau)=\sum_{\yy\ne\xx}\Big(p_{x}R_{\yx}\ln\frac{p_{x}R_{\yx}}{p_{y}R_{\oddconj x\oddconj y}}-p_{x}R_{\yx}+p_{y}R_{\oddconj x\oddconj y}\Big).\label{eq:eprodd-1}
\end{align}
Note that, in general, \cref{eq:eprodd} does not have the usual ``flux-force''
form $\epr=\sum_{\yy\ne\xx}p_{x}R_{\yx}\ln\frac{p_{x}R_{\yx}}{p_{y}R_{\oddconj x\oddconj y}}$,
as it does in systems without odd variables.

In deriving \cref{eq:eprodd-1}, we assumed that the system is coupled
to a single heat bath. However, the derivation can be generalized
to multiple heat baths (or other types of reservoirs), as often done
in stochastic thermodynamics \citep{esposito2010threefaces}. Let
$T_{y,\bath\vert x}(\tau)$ indicate the conditional probability that
the system is in state $y$ and last exchanged energy with bath $\bath$
at time $t+\tau$, given that the system was in state $x$ at time
$t$ ($\alpha=\varnothing$ when $x=y$). Assuming $T_{y,\bath\vert x}$
arises from the continuous-time generator $R^{\bath}$, we can then
generalize \cref{eq:eprodd-1} to
\begin{align}
\epr=\sum_{\yy\ne\xx,\bath}\Big(p_{x}R_{\yx}^{\bath}\ln\frac{p_{x}R_{\yx}^{\bath}}{p_{y}R_{\oddconj x\oddconj y}^{\bath}}-p_{x}R_{\yx}^{\bath}+p_{y}R_{\oddconj x\oddconj y}^{\bath}\Big).\label{eq:eprodd}
\end{align}

\subsection{EPR as a generalized KL divergence}
 
We now show that the EPR in a system with odd variables, as derived
in \cref{eq:eprodd}, can be expressed in our formalism as a generalized
KL divergence between elements of an exponential  family.

As for stochastic master equations without odd variables, we define one reaction $\rr$ for each transition $(x\to y,\bath)$ ($x\ne y$), whose reverse reaction $\negedge$ corresponds to the transition $(y\to x,\bath)$. The fluxes of these two reactions are given by  $\jjrr=p_{x}R_{\yx}^{\bath}$ and $\jjEdgeNegRev = p_{y} R_{\xy}^{\bath}$, as usual. Next, we
define the thermodynamic force across reaction $\rr$ as $\ffrr=\ln[(p_{x}R_{\yx}^{\bath})/(p_{y}R_{\oddconj x\oddconj y}^{\bath})]$, in line with \cref{eq:conjldb}. 
We emphasize that in the presence of odd variables, in general $\ffrr \ne \ln(\jjrr/\jjEdgeNegRev)$. 
Next, we define the exponential  family of fluxes $\jjrrT{\zz}=\jjrr e^{\eerr -\ffrr}$, exactly as in  \mainref{eq:pfam}. For the reaction $\rr$ corresponding to $(x,y,\alpha)$, $\jjrrT{\ff}=\jjrr$ and $\jjrrT{\zz} = \jjrr e^{-\ffrr}=p_{y}R_{\oddconj x\oddconj y}^{\bath}$. 

\cref{eq:eprodd} can
then be written as the generalized KL divergence between $\jjT{\ff}$ and $\jjT{\zz}$,
\begin{align}
\epr=\sum_{\rr}\Big(\jjrr\ln\frac{\jjrr }{\jjrr e^{-\ffrr}}-\jjrr+\jjrr e^{-\ffrr}\Big)=\DD(\ff\Vert\zz),\label{eq:eprodd2}
\end{align}
as in \mainref{eq:eprdist} in the main text. 
Our housekeeping/excess decomposition of EPR --- as described in
\mainref{eq:hkdef}, \mainref{eq:maxent}, and \mainref{eq:pyth}
--- depends only on the fact that the EPR can be expressed as $\epr=\DD(\ff\Vert\zz)$.
Therefore, those definitions apply without modification to systems
with odd variables.

As we note in the main text, some of the subsequent results do depend
on properties of systems without odd variables. For instance, some
of our results exploit the symmetry $\BBdiv\jj=-\BBdiv\jjT{\zz}$ from \cref{eq:app-reverse}, which
in general won't hold for systems with odd variables. For instance,
for systems with odd variables, it is no longer guaranteed that excess
EPR vanishes in steady state. However, it will vanish as long as an
additional condition is satisfied, which is that the steady state
is symmetric under conjugation odd variables, $\pi_{x}=\pi_{\oddconj x}$.

This is related to the fact that, in systems with odd variables, the steady state may be
out of equilibrium even when the thermodynamic
forces are conservative. However, the steady state will always be
in equilibrium if the forces are conservative and the steady-state
distribution is symmetric under conjugation of odd variables. 
See Ref.~\citep{lee2013fluctuation} for further discussion.

See Section~\oldref{app:hs} for a comparison of our decomposition
and the HS decomposition in systems with odd variables.

\subsection{Example: particle on a ring}

\label{app:odd-example}

We now provide a simple example to illustrate our decomposition on
a system with odd variables. It will be shown that our excess and
housekeeping EPR terms are always nonnegative, unlike the HS decomposition
where the housekeeping EPR can take unphysical negative values~\citep{lee2013fluctuation,spinney2012nonequilibrium,ford2012entropy}.

We use a standard model from the literature on the stochastic thermodynamics of systems 
with odd variables \citep{lee2013fluctuation,spinney2012nonequilibrium,ford2012entropy}, illustrated in \cref{fig:oddexample}.
There is a particle on a ring with $L$ locations, which has a binary
velocity degree of freedom which is odd. Formally, the system's state
is given by $x=(r,v)$, where $r\in\{1,\dots,L\}$ is the position
of the particle on the ring and $v\in\{{-1},+1\}$ is the velocity.
For $x=(r,v)$, the conjugated state is given by $\oddconj x=(r,-v)$.
The particle moves in the direction of its velocity, $(r,v)\to(r+v,v)$,
with rate $e^{\alpha}$ when $v=+1$ and rate $1$ when $v=-1$. In
addition, the velocity flips as $(r,v)\to(r,-v)$ with rate $e^{\gamma}$
when $v=+1$ and rate 1 when $v=-1$. The thermodynamic forces 
across the two types of transitions are
\begin{alignat*}{2}
f_{(r,v)\to(r+v,v)}&=\begin{cases}
\ln\frac{p_{r,v}e^\alpha}{p_{r+v,v}} & v=+1\\
\ln\frac{p_{r,v}}{p_{r+v,v}e^\alpha} & v=-1
\end{cases}
&&= \ln p_{r,v}-\ln p_{r+v,v}+v\alpha,
\\
f_{(r,v)\to(r,-v)}&=
\begin{cases}
\ln\frac{p_{r,v}e^\gamma}{p_{r,-v}e^\gamma} & v=+1\\
\ln\frac{p_{r,v}}{p_{r,-v}} & v=-1
\end{cases}
&&=\ln p_{r,v}-\ln p_{r,-v}. 
\end{alignat*}
The steady-state distribution is given by 
\begin{align}
\pi_{r,v}=\frac{\delta_{v,1}+\delta_{v,-1}e^{\gamma}}{L(e^{\gamma}+1)}.\label{eq:odd-ss}
\end{align}
In this model, the parameter $\alpha$ controls the breaking of symmetry
for the two direction of movement around the ring, leading to nonconservative
forces when $\alpha\ne0$. The parameter $\gamma$ controls the breaking
of symmetry of velocity flips, leading to a steady-state distribution that is not
symmetric under conjugation of odd variables ($\pi_{r,v}\ne\pi_{r,-v}$)
when $\gamma\ne0$. The steady state is in equilibrium,  only when
$\alpha=0$ and $\gamma=0$.

\begin{figure}[h]
    \centering
    \includegraphics[width=.5\linewidth]{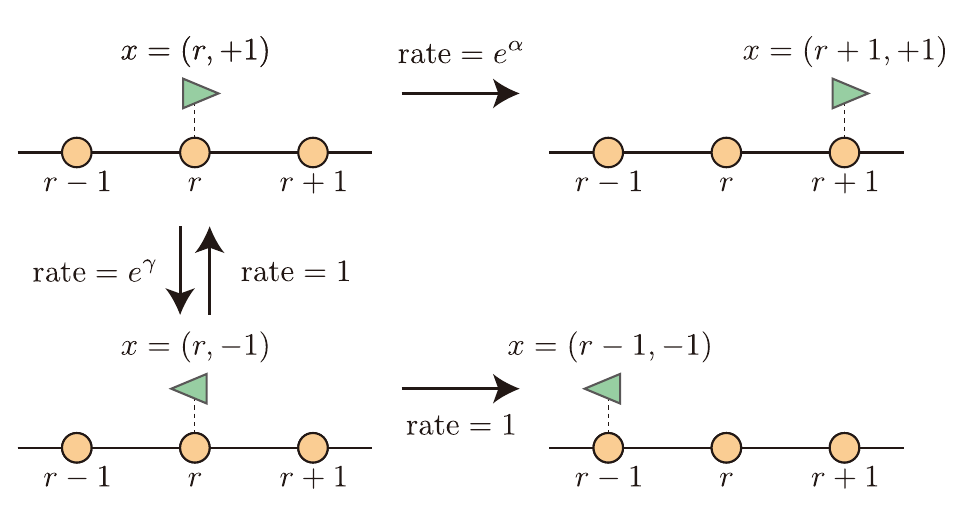}
    \caption{We consider a standard model of a discrete system with an odd degrees of freedom~\cite{lee2013fluctuation,spinney2012nonequilibrium,ford2012entropy}: a particle on a ring of $L$ states with position $r\in \{1,\dots,L\}$ and velocity $v\in\{-1,+1\}$, where the velocity is odd.}
    \label{fig:oddexample}
\end{figure}

In \cref{fig:odd}, we visualize the time-dependent values of EPR $\epr$,
our excess EPR $\eprEX$, and the HS excess EPR $\eprhsEX$. We consider
a system with $L=4$ positions and the non-stationary initial distribution
$p_{rv}\propto10\delta_{r0}\delta_{v1}+1$. We consider four conditions:
\begin{enumerate}
\item $\alpha=0$ and $\gamma=0$, so that all transitions are symmetric.
Here the forces are conservative, $f=\negBBgrad\pot$ for $\phi_{x}=\ln p_{x}$
and the steady-state distribution is symmetric under conjugation of odd variables.
The steady state is in equilibrium and $\epr=\eprEX=\eprhsEX$ at
all times.
\item $\alpha=0$ and $\gamma=1$, so velocity flips $(r,-1)\to(r,+1)$ occur
more frequently than $(r,+1)\to(r,-1)$. The forces are conservative,
$f=\negBBgrad\pot$ for $\phi_{x}=\ln p_{x}$, but the steady state
is not symmetric under conjugation of odd variables. The steady state
is not in equilibrium ($\epr>0$ in steady state). Since the forces
are conservative, under our decomposition the housekeeping EPR vanishes
and $\epr=\eprEX$ at all times. The HS decomposition gives different
results, which can take unphysical negative values: $\eprhsEX>\epr,\eprhsHK<0$.
\item $\alpha=1$ and $\gamma=0$, so movements along the ring with positive
velocity are faster than those with negative velocity. The steady
state is symmetric under time-reversal but the forces $f_{(r,v)\to(r+v,v)}$
are not conservative, so the steady state is out of equilibrium. Our
decomposition and HS decomposition both obey $0\le\eprEX\le\epr$
and $0\le\eprhsEX\le\epr$. We also verify that, in systems with steady
states that symmetric under time reversal, $\eprEX\ge\eprhsEX$ always
(\cref{eq:appbndHS} in \cref{app:hs}) and $\eprEX=\eprhsEX=0$ in
steady state.
\item $\alpha=1$ and $\gamma=1$, so the forces are not conservative and
the steady state is not symmetric under conjugation of odd variables.
The HS decomposition again gives unphysical values $\eprhsEX>\epr,\eprhsHK<0$.
Under our decomposition, neither $\eprEX$ nor $\eprHK$ vanish in
steady state.
\end{enumerate}
\vspace{15pt}

\begin{figure}[H]
\centering \includegraphics[width=1\textwidth]{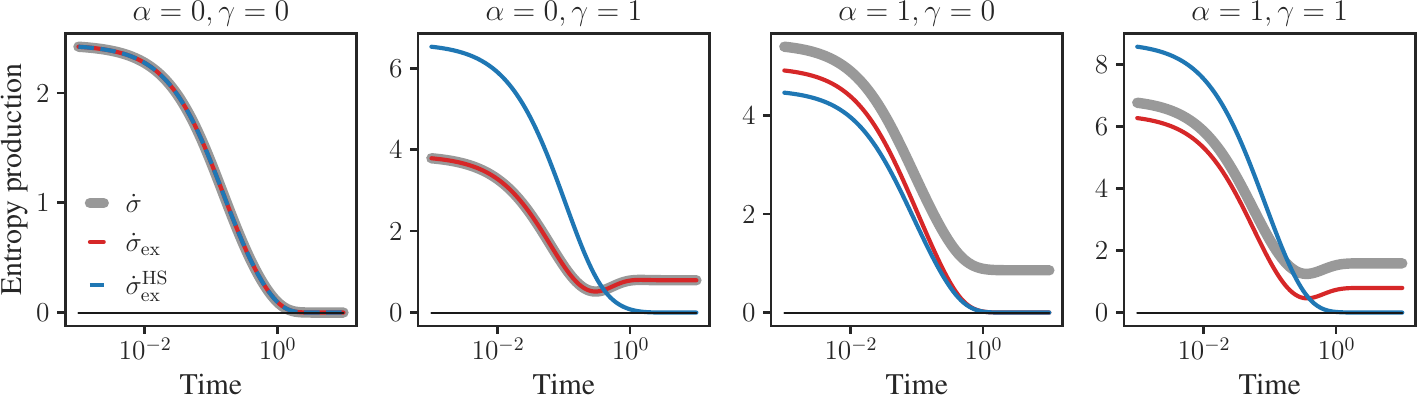} \caption{Plots of overall EPR $\protect\epr$, our excess EPR $\protect\eprEX$,
and the HS excess EPR $\protect\eprhsEX$ for the ring model with
an odd velocity variable. When $\alpha\protect\ne0$ the forces are
non-conservative, when $\gamma\protect\ne0$ the steady-state distribution is not
symmetric under conjugation of odd variables. The HS decomposition
can give unphysical values ($\protect\eprhsEX>\protect\epr,\protect\eprhsHK<0$)
when $\gamma\protect\ne0$.}
\label{fig:odd}
\end{figure}

\section{Comparisons with other decompositions}

\subsection{Hatano-Sasa decomposition}

\noindent \label{app:hs}
\global\long\def\complexes{C}%
\global\long\def\ssfluxrr{\mathcal{J}_{\rr}^{\sss}}%
Here we compare our information-geometric housekeeping/excess decomposition,
$\epr=\eprEX+\eprHK$, to the HS housekeeping/excess decomposition,
$\epr=\eprhsHK+\eprhsEX$. We derive the following inequality: 
\begin{align}
\eprHK\le\eprhsHK\qquad\eprEX\ge\eprhsEX\label{eq:appbndHS}
\end{align}
for (1) stochastic master equations without odd variables, (2) stochastic
master equations with odd variables and time-symmetric steady states,
and (3) chemical systems with complex balance and mass action kinetics.
In all cases, we show that the HS housekeeping EPR can be written
as the generalized KL divergence
\begin{align}
\eprhsHK=\DD(\ff\Vert\negBBgrad\pot^{\sss}),\label{eq:df2}
\end{align}
where $\phi_{x}^{\sss}:=\ln(p_{x}/\pi_{x})$ is defined via the steady-state
distribution $\bm{\pi}$. Since our housekeeping EPR satisfies the
variational principle in \mainref{eq:hkdef}, \cref{eq:df2} implies
\cref{eq:appbndHS}.

We first consider the simplest case, a stochastic master equation
without odd variables. The EPR is given by \citep{esposito2010threefaces}
\begin{align}
\epr=\sum_{\yy\ne\xx,\bath}p_{\xx}R_{\yx}^{\bath}\ln\frac{p_{\xx}R_{\yx}^{\bath}}{p_{\yy}R_{\xx\yy}^{\bath}},
\end{align}
where $\bath$ indexes over thermodynamic reservoirs. The HS excess
and housekeeping terms are then given by~\citep{esposito2007entropy,esposito2010threefaces}
\begin{align}
\eprhsEX & =\sum_{\yy\ne\xx,\bath}p_{\xx}R_{\yx}^{\bath}\ln\frac{p_{x}\pi_{\yy}}{\pi_{\xx}p_{\yy}}\label{eq:hs01}\\
\eprhsHK & =\epr-\eprhsEX=\sum_{\yy\ne\xx,\bath}p_{\xx}R_{\yx}^{\bath}\ln\frac{p_{\xx}R_{\yx}^{\bath}}{p_{\xx}R_{\oddconj\xx\oddconj\yy}^{\bath}\pi_{\yy}/\pi_{\xx}}.\label{eq:hs01b}
\end{align}
Within our exponential family \mainref{eq:pfam}, the potential $\pot^{\mathrm{ss}}$
gives rise to the fluxes 
\[
[\jjT{\negBBgrad\pot^{\mathrm{ss}}}]_{x\to y;\bath}=p_{y}R_{\oddconj x\oddconj y}^{\bath}e^{\ln p_{\xx}/\pi_{\xx}-\ln p_{y}/\pi_{y}}=p_{x}R_{\oddconj x\oddconj y}^{\bath}\pi_{y}/\pi_{x}.
\]
It leads to the following generalized KL divergence,
\begin{align}
\DD(\ff\Vert\negBBgrad\pot^{\sss})= & \sum_{\yy\ne\xx,\bath}p_{\xx}R_{\yx}^{\bath}\ln\frac{p_{\xx}R_{\yx}^{\bath}}{p_{x}R_{\oddconj x\oddconj y}^{\bath}\pi_{\yy}/\pi_{\xx}}-p_{\xx}R_{\yx}^{\bath}+p_{x}R_{\oddconj x\oddconj y}^{\bath}\frac{\pi_{y}}{\pi_{x}}\label{eq:df3}\\
= & \sum_{\yy\ne\xx,\bath}p_{\xx}R_{\yx}^{\bath}\ln\frac{p_{\xx}R_{\yx}^{\bath}}{p_{x}R_{\oddconj x\oddconj y}^{\bath}\pi_{\yy}/\pi_{\xx}}=\eprhsHK,\label{eq:ddd}
\end{align}
where in the second line we used that 
\begin{align*}
\sum_{y\ne x,\bath}\left(p_{\xx}R_{\oddconj\xx\oddconj\yy}^{\bath}\frac{\pi_{y}}{\pi_{x}}-p_{\xx}R_{\yx}^{\bath}\right) & =\sum_{x}\frac{p_{\xx}}{\pi_{\xx}}\sum_{y(\ne x),\bath}\left(\pi_{\yy}R_{\oddconj\xx\oddconj\yy}^{\bath}-\pi_{\xx}R_{\yx}^{\bath}\right)=0,
\end{align*}
which follows since $\bm{\pi}$ is a steady-state distribution. In
this way, we derived \cref{eq:df2} in \cref{eq:ddd}.

Next, we consider stochastic master equations with odd variables,
under the assumption that the steady-state distribution is symmetric
under conjugation of odd variables $\pi_{x}=\pi_{\oddconj x}$. As
shown in \cref{eq:eprodd} in \cref{app:odd}, the EPR is
\begin{align}
\epr=\sum_{\yy\ne\xx,\bath}\big(p_{\xx}R_{\yx}^{\bath}\ln\frac{p_{\xx}R_{\yx}^{\bath}}{p_{\yy}R_{\oddconj\xx\oddconj\yy}^{\bath}}-p_{\xx}R_{\yx}^{\bath}+p_{\yy}R_{\oddconj\xx\oddconj\yy}^{\bath}\big).
\end{align}
The HS excess EPR is still defined as in \cref{eq:hs01}, while the
HS housekeeping EPR is the remainder \citep{spinney2012nonequilibrium,lee2013fluctuation}:
\begin{align}
\eprhsHK & =\epr-\eprhsEX=\sum_{\yy\ne\xx,\bath}\big(p_{\xx}R_{\yx}^{\bath}\ln\frac{p_{x}R_{\yx}^{\bath}}{p_{x}R_{\oddconj\xx\oddconj\yy}^{\bath}\pi_{\yy}/\pi_{\xx}}-p_{\xx}R_{\yx}^{\bath}+p_{\yy}R_{\oddconj\xx\oddconj\yy}^{\bath}\big).%
\label{eq:hs02-2}
\end{align}
The potential $\pot^{\mathrm{ss}}$ gives rise to the fluxes 
\[
[\jjT{\negBBgrad\pot^{\mathrm{ss}}}]_{x\to y;\bath}=p_{y}R_{\oddconj\xx\oddconj\yy}^{\bath}e^{\ln p_{\xx}/\pi_{\xx}-\ln p_{y}/\pi_{y}}=p_{x}R_{\oddconj\xx\oddconj\yy}^{\bath}\frac{\pi_{y}}{\pi_{x}},
\]
and leads to the following generalized KL divergence
\begin{align*}
\DD(\ff\Vert\negBBgrad\pot^{\sss})= & \sum_{\yy\ne\xx,\bath}p_{\xx}R_{\yx}^{\bath}\ln\frac{p_{\xx}R_{\yx}^{\bath}}{p_{x}R_{\oddconj\xx\oddconj\yy}^{\bath}\frac{\pi_{y}}{\pi_{x}}}-p_{\xx}R_{\yx}^{\bath}+p_{x}R_{\oddconj\xx\oddconj\yy}^{\bath}\frac{\pi_{y}}{\pi_{x}}.
\end{align*}
Finally, we have 
\begin{align*}
\sum_{y\ne x,\bath}\Big(p_{\xx}R_{\oddconj\xx\oddconj\yy}^{\bath}\pi_{\yy}/\pi_{\xx}-p_{\yy}R_{\oddconj\xx\oddconj\yy}^{\bath}\Big) & =\sum_{y\ne x,\bath}\Big(p_{\xx}R_{\oddconj\xx\oddconj\yy}^{\bath}\pi_{\yy}/\pi_{\xx}-p_{x}R_{\oddconj y\oddconj x}^{\bath}\Big)\\
 & =\sum_{x}\frac{p_{\xx}}{\pi_{\xx}}\sum_{y(\ne x),\bath}\Big(R_{\oddconj\xx\oddconj\yy}^{\bath}\pi_{\yy}-R_{\oddconj\xx\oddconj\yy}^{\bath}\pi_{\xx}\Big)\\
 & =\sum_{x}\frac{p_{\xx}}{\pi_{\xx}}\sum_{y(\ne x),\bath}\Big(R_{\oddconj\xx\oddconj\yy}^{\bath}\pi_{\oddconj\yy}-R_{\oddconj\xx\oddconj\yy}^{\bath}\pi_{\oddconj\xx}\Big)=0,
\end{align*}
where we used the symmetry $\pi_{\xx}=\pi_{\oddconj\xx}$. Plugging
$\sum_{y\ne x,\bath}p_{\xx}R_{\oddconj\xx\oddconj\yy}^{\bath}\frac{\pi_{\yy}}{\pi_{\xx}}=\sum_{y\ne x,\bath}p_{\yy}R_{\oddconj\xx\oddconj\yy}^{\bath}$
into \cref{eq:hs02-2} implies \cref{eq:df2}.

\global\long\def\ww{\bm{\eta}}%
\global\long\def\wwxx{{\eta}_{\xx}}%

Finally, we consider chemical systems that obey complex balance,
meaning that the net current entering and leaving each chemical complex
vanishes in steady state~\citep{feinbergFoundationsChemicalReaction2019}.
We also assume mass action kinetics, as in \cref{eq:massaction}. In
that case, the HS excess and housekeeping EPR is~\citep{ge2016nonequilibrium,rao2016nonequilibrium}
\begin{align}
\eprhsEX & =-\sum_{\rr}\jjrr\sum_{\xx}\BBxxrr\ln\frac{c_{x}}{\pi_{\xx}}.\label{eq:hs03}\\
\eprhsHK & =\epr-\eprhsEX=\sum_{\rr}\jjrr\left[\ln\frac{\jjrr}{\jjEdgeNegRev}+\sum_{\xx}\BBxxrr\ln\frac{c_{x}}{\pi_{\xx}}\right],\label{eq:hs04}
\end{align}
where $\cc$ is the vector actual concentrations and $\bm{\pi}$ is
the vector of steady-state concentrations ($\cc$ and $\bm{\pi}$
are nonnegative, but do not necessarily sum to 1). Using the potential
$\phi_{\xx}^{\text{ss}}:=\ln(c_{\xx}/\pi_{\xx})$, \cref{eq:hs04}
can be written as 
\begin{align*}
 & \eprhsHK=\jj^{T}(\ff+\BBgrad\pot^{\sss}).
\end{align*}
Using this expression and \mainref{eq:klgen}, we write 
\begin{align}
\DD(\ff\Vert\negBBgrad\pot^{\sss})=\eprhsHK-V,\label{eq:appg5}
\end{align}
where for notational convenience we defined
\begin{align}
V=\sum_{\rr}(\jjrr-\jjEdgeNegRev e^{[\negBBgrad\pot^{\sss}]_{\rr}}).\label{eq:vv}
\end{align}
We prove \cref{eq:df2} by showing that $V=0$.

To begin, split the right hand side of \cref{eq:vv} into contributions
from the forward and negative side of each reversible reaction $r$
(see discussion of notation in \cref{app:graph}), 
\begin{align}
V=\sum_{r}(\jjrrCRN-\jjRevrrCRN e^{[\negBBgrad\pot^{\sss}]_{r}})+\sum_{r}(\jjrrCRN-\jjRevrrCRN e^{[{\BBgrad}\pot^{\sss}]_{r}}).\label{eq:ap2s}
\end{align}
Using \cref{eq:massaction}, each term in the first sum can be written
as 
\begin{align*}
\jjrrCRN-\jjRevrrCRN e^{[\negBBgrad\pot^{\sss}]_{r}} & =k_{r}^{\rightarrow}\prod_{\xx}c_{\xx}^{\nu_{\xx r}}-k_{r}^{\leftarrow}\prod_{\xx}c_{\xx}^{\kappa_{\xx r}}\prod_{\xx}\Big(\frac{c_{\xx}}{\pi_{\xx}}\Big)^{\nu_{\xx r}-\kappa_{\xx r}}\\
 & =\prod_{\xx}\Big(\frac{c_{\xx}}{\pi_{\xx}}\Big)^{\nu_{\xx r}}\left(k_{r}^{\rightarrow}\prod_{\xx}\pi_{\xx}^{\nu_{\xx r}}-k_{r}^{\leftarrow}\prod_{\xx}\pi_{\xx}^{\kappa_{\xx r}}\right)=\prod_{\xx}\Big(\frac{c_{\xx}}{\pi_{\xx}}\Big)^{\nu_{\xx r}}\ssfluxrr,
\end{align*}
where $\ssfluxrr$ is the current (net flux) across reversible reaction
$r$ in steady state. Using a similar derivation, we write each term
in the second sum in \cref{eq:ap2s} as 
\begin{align}
 & \jjrrCRN-\jjRevrrCRN e^{[\BBgrad\pot^{\sss}]_{r}}=-\prod_{\xx}\Big(\frac{c_{\xx}}{\pi_{\xx}}\Big)^{\kappa_{\xx r}}\ssfluxrr.
\end{align}
Combining, we rewrite \cref{eq:vv} as 
\begin{align}
V=\sum_{r}\left[\prod_{\xx}\Big(\frac{c_{\xx}}{\pi_{\xx}}\Big)^{\nu_{\xx r}}\ssfluxrr-\prod_{\xx}\Big(\frac{c_{\xx}}{\pi_{\xx}}\Big)^{\kappa_{\xx r}}\ssfluxrr\right].\label{eq:s2}
\end{align}
Now split the right hand side into contributions from each reactant
complex and each product complex. Let $\complexes$ indicate the set
of reactant and product complexes, where each element of $\complexes$
is a vector $\ww\in\mathbb{\mathbb{N}}_{0}^{\numstate}$ with $\wwxx$
is the number of species $\xx$ in complex $\ww$. Let $A(\ww)=\{r:\nu_{xr}=\wwxx\forall x\}$
indicate the set of reactions that have reactant complex $\ww$, and
let $B(\ww)=\{r:\kappa_{xr}=\wwxx\forall x\}$ indicate the set of
reactions that have product complex $\ww$. Then, we can rewrite  \cref{eq:s2} 
\begin{align*}
V&=\sum_{\ww\in\complexes}\left[\sum_{r\in A(\ww)}\prod_{\xx}\Big(\frac{c_{\xx}}{\pi_{\xx}}\Big)^{\nu_{\xx r}}\ssfluxrr-\sum_{r\in B(\ww)}\prod_{\xx}\Big(\frac{c_{\xx}}{\pi_{\xx}}\Big)^{\kappa_{\xx r}}\ssfluxrr\right] \\
& =\sum_{\ww\in\complexes}\left[\sum_{r\in A(\ww)}\prod_{\xx}\Big(\frac{c_{\xx}}{\pi_{\xx}}\Big)^{\wwxx}\ssfluxrr-\sum_{r\in B(\ww)}\prod_{\xx}\Big(\frac{c_{\xx}}{\pi_{\xx}}\Big)^{\wwxx}\ssfluxrr\right]\\
 & =\sum_{\ww\in\complexes}\prod_{\xx}\Big(\frac{c_{\xx}}{\pi_{\xx}}\Big)^{\wwxx}\left[\sum_{r\in A(\ww)}\ssfluxrr-\sum_{r\in B(\ww)}\ssfluxrr\right].
\end{align*}
By the definition of complex balance, $\sum_{r\in A(\ww)}\ssfluxrr=\sum_{r\in B(\ww)}\ssfluxrr$
for each $\ww$ ~\citep{feinbergFoundationsChemicalReaction2019}.
Therefore, $V=0$, which implies \cref{eq:df2}.

\subsection{``Onsager-projective decomposition'' from Ref.~\NoCaseChange{\citep{kohei2022}}}

\label{app:ons}

This paper builds on recent work by the present authors~\citep{kohei2022},
which studied excess and housekeeping EPR in discrete Markovian
systems. It considered both on linear stochastic master equations and nonlinear
chemical reaction networks, though only without odd variables (see also Refs.~\citep{dechant2022geometric,dechant2022geometricCoupling}
for continuous systems). 

As in the present paper, Ref.~\citep{kohei2022}
considers the excess and housekeeping decomposition from a geometric
perspective. In that paper, the EPR is written as the squared (generalized)
Euclidean norm of the force vector under an appropriate metric: 
\begin{align}
\epr=\left\Vert \ff\right\Vert _{L}^{2}\equiv\ff^{T}L\ff,
\end{align}
where $\ff\in\mathbb{R}^{\numedge}$ is the thermodynamic force (same
as in this paper) and $L$ is a diagonal matrix $\mathbb{R}_{+}^{\numedge\times\numedge}$
of edgewise Onsager coefficients,
\begin{align}
L_{\rr\rr}=\frac{1}{2}(\jjrr-\jjEdgeNegRev)/\ffrr.\label{eq:d1-1}
\end{align}
(The factor 1/2 appears here but not in Ref.~\citep{kohei2022} due
to a minor change of convention: unlike Ref.~\citep{kohei2022},
in this paper we consider reversible reactions as two separate reactions.)
The force vector is projected onto the subspace of conservative forces,
which gives rise to the optimal potential: 
\begin{align}
\pot_{\text{ons}}^{*}=\argmin_{\pot\in\mathbb{R}^{\numedge}}\big\Vert\ff-(\negBBgrad\pot)\big\Vert_{L}^{2}.\label{eq:proj1}
\end{align}
where the subscript ``ons'' refers to the Onsager metric. The housekeeping
EPR is then defined as the squared (generalized) Euclidean distance
from $\ff$ to the subspace of conservative forces, while the excess
EPR is defined as the squared (generalized) Euclidean norm of the
projected conservative force, 
\begin{align}
\epr=\left\Vert \ff\right\Vert _{L}^{2}=\underbrace{\left\Vert \ff-(\negBBgrad\pot_{\text{ons}}^{*})\right\Vert _{L}^{2}}_{\eprHKons}+\underbrace{\left\Vert \BBgrad\pot_{\text{ons}}^{*}\right\Vert _{L}^{2}}_{\eprEXons}.\label{eq:pythons}
\end{align}
We refer to $\eprHKons$ and $\eprEXons$ as the \emph{Onsager-projective}
housekeeping and excess EPR terms.

In this paper, we work within the non-Euclidean setting of information
geometry. In our case, distance is measured in terms of KL divergence
rather than generalized Euclidean norm. Nonetheless, it is clear that
\mainref{eq:hkdef} is the information-geometric analogue of \cref{eq:proj1},
while \mainref{eq:pyth} is the information-geometric analogue of
\cref{eq:pythons}. Thus, our approach is an information-geometric
extension of Ref.~\citep{kohei2022}.

Euclidean geometry suffices for systems that exhibit Onsager-type
linear relations between thermodynamic forces and fluxes, as occurs
near steady state or near equilibrium. On the other hand, far-from-equilibrium
analysis requires an information-geometric treatment. For this reason,
the TURs and {TSLs} derived in Ref.~\citep{kohei2022} are in general
only tight for systems that are close to equilibrium and/or steady
state, while the bounds derived in this paper can be tight arbitrarily
far from equilibrium. %

However, we can relate the two decompositions. In accordance with
Ref.~\citep{kohei2022}, we restrict our attention to systems without
odd variables, and show that 
\begin{align}
\eprHK\ge\eprHKons\qquad\eprEX\le\eprEXons.\label{eq:onsineq}
\end{align}
To derive this result, recall that for systems without odd variables, 
each reaction $\rr$ is paired with a unique reverse reaction $\negedge$ 
such that $\ffrrR=-\ffrr$. Consider the
KL divergence between the forward fluxes $\jj=\jjT{\ff}$ and any
other $\jjT{\ee}$ where $\ee$ is anti-symmetric ($\eerr=-\eerrR$):
\begin{align*}
\DD(\ff\Vert\ee) & =\sum_{\rr}\jjrr(e^{-(\ffrr-\eerr)}+(\ffrr-\eerr)-1)\\
 & =\sum_{\rr}\jjEdgeNegRev(e^{\ffrr-\eerr}-(\ffrr-\eerr)-1).
\end{align*}
On the first line we rearranged \mainref{eq:klgen} in the main text, and in the second
line we used anti-symmetry of $\ff$ and $\ee$. Combining these expressions,
and using $\jjrr=e^{\ffrr}\jjEdgeNegRev$, gives 
\[
\DD(\ff\Vert\ee)=\frac{1}{2}\sum_{\rr}\jjEdgeNegRev\big((e^{\ffrr-\eerr}-(\ffrr-\eerr)-1)+e^{\ffrr}(e^{-(\ffrr-\eerr)}+(\ffrr-\eerr)-1)\big).
\]
We now rewrite the right hand side as 
\begin{align}
\DD(\ff\Vert\ee) & =\frac{1}{2}\sum_{\rr}\jjEdgeNegRev\big(h(\ffrr-\eerr,\ffrr)+\frac{e^{f}-1}{f}(\ffrr-\eerr)^{2}\big)\nonumber \\
 & =\frac{1}{2}\sum_{\rr}\jjEdgeNegRev h(\ffrr-\eerr,\ffrr)+\left\Vert \ff-\ee\right\Vert _{L}^{2},\!\!\!\!\!\!\label{eq:d1}
\end{align}
where for convenience we defined the following function: 
\[
h(a,b)=\Big[\frac{(e^{a}-a-1)+e^{b}(e^{-a}+a-1)}{a^{2}}-\frac{e^{b}-1}{b}\Big]a^{2}.
\]
It can be verified (e.g., by taking derivatives with respect to $a$)
that the term inside the brackets is nonnegative. Thus, $h$ is nonnegative,
therefore $\DD(\ff\Vert\ee)\ge\left\Vert \ff-\ee\right\Vert _{L}^{2}$
given \cref{eq:d1}. Finally, since $\ee=\negBBgrad\pot$ is anti-symmetric,
we arrive at \cref{eq:onsineq}: 
\[
\eprHK=\min_{\pot}\DD(\ff\Vert\negBBgrad\pot)\ge\min_{\pot}\left\Vert \ff-(\negBBgrad\pot)\right\Vert _{L}^{2}=\eprHKons.
\]
For a numerical comparison between the decomposition proposed in this
paper and Ref.~\citep{kohei2022}, see \cref{app:numericalcomp}.

We now consider the limit in which the two decompositions agree. Using
the derivations above, we have the bounds 
\begin{align}
0\le\eprHK-\eprHKons\le\frac{1}{2}\sum_{\rr}\jjEdgeNegRev\,h(\ffrr+[\BBgrad\pot_{\text{ons}}^{*}]_{\rr},\ffrr).\label{eq:d2-1}
\end{align}
The function $h(a,b)$ vanishes to first order around $a=b$ and $a=0$
(in general, $h(a,b)$ is symmetric under the transformation $a\mapsto b-a$).
In the context of \cref{eq:d2-1}, $a=b$ reflects that $\eprHK$ and
$\eprHKons$ agree to first order around $\BBgrad\pot_{\text{ons}}^{*}=0$
(steady state) while $a=0$ reflects that they agree to first order
around $\ff=\negBBgrad\pot_{\text{ons}}^{*}$ (the forces are conservative).
We can ask if they also agree to second order there. A Taylor expansion
of $h(a,b)$ shows that second order terms do not vanish except in
the limit $b\to0$. In the context of \cref{eq:d2-1}, this is the
equilibrium limit $f_{\rr}\to0$, where the thermodynamic force across
each reaction vanishes.  Note that 
\[
c_{1}\left\Vert \ff\right\Vert ^{2}\ge\left\Vert \ff\right\Vert _{L}^{2}\ge\left\Vert \ff+\BBgrad\pot_{\text{ons}}^{*}\right\Vert _{L}^{2}\ge c_{2}\left\Vert \ff+\BBgrad\pot_{\text{ons}}^{*}\right\Vert ^{2}
\]
where $c_{1}=\max(\jjrr+\jjEdgeNegRev)/2$, $c_{2}=\min_{\rr}\sqrt{\jjrr\jjEdgeNegRev}$,
and $\Vert\cdot\Vert$ is the usual Euclidean norm (we used \cref{eq:pythons}
for the middle inequality, the others come from bounds on the logarithmic
mean in \cref{eq:d1-1}). Thus, if $f_{\rr}\to0$, we can assume that
$\ffrr+[\BBgrad\pot_{\text{ons}}^{*}]_{\rr}$, first argument of $h$,
also vanishes in \cref{eq:d2-1}. We now expand $h$ in each argument
and rearrange to give 
\begin{align}
h(\gamma,f)=\frac{1}{12}\gamma^{2}\left(\gamma-f\right)^{2}+\mathcal{O}(\epsilon^{5})=\mathcal{O}(\epsilon^{4})
\end{align}
for $f,\gamma\sim\epsilon$. Plugging into \cref{eq:d2-1} shows that
$\eprHK$ and $\eprHKons$ agree to third order in the equilibrium
limit.

As discussed in Ref.~\citep{kohei2022}, the Onsager-projective decomposition
can be seen as an extension of the Maes and Netočný (MN) approach~\citep{maes2014nonequilibrium}
to discrete systems. The decomposition proposed in this paper can
be seen as a generalization of the MN decomposition to the far-from-equilibrium
regime.

\subsection{Numerical comparison with Refs.~\NoCaseChange{\citep{kohei2022}}
and~\NoCaseChange{\citep{Kobayashi2022}}}

\label{app:numericalcomp}
\global\long\def\netforce{\bm{\mathcal{F}}}%
\global\long\def\netforcerr{\mathcal{F}_{r}}%
Here, we numerically compare three decompositions: the Onsager-projective
decomposition described in \cref{app:ons}, the ``Hessian decomposition''
which was recently proposed in Ref.~\citep{Kobayashi2022}, and the
information-geometric decomposition that we propose in this Letter.
While an inequality exists between the Onsager decomposition and our
decomposition, \cref{eq:onsineq}, no inequality between the Hessian
decomposition and the others has been proved analytically. Nonetheless,
our numerical results prove that they are different. They also suggest
that the Hessian decomposition gives intermediate values between the
other two decompositions.

To be self-contained, we briefly review the Hessian decomposition
presented in Ref.~\citep{Kobayashi2022}. Consider a system without
odd variables that has $\numedge$ reactions, having forward and reverse fluxes 
$\jjrr$ and $\jjEdgeNegRev$. We adopt the notation defined in \cref{app:graph}:
we use $r\in\{1,2,\dots,M/2\}$ to label each pair of reactions $\rr$
and $\negedge$, where the forward/reverse fluxes of the pair are indicated
as $\jjrrCRN=\jjrr$ and $\jjRevrrCRN=\jjEdgeNegRev$. We define
a vector of currents (net fluxes) $\netflux\in\mathbb{R}^{M/2}$ as
$\netjR:=\jjrrCRN-\jjRevrrCRN$, a vector of ``frenetic activities''
$\bm{\omega}\in\mathbb{R}_{+}^{M/2}$ as $\omega_{r}:=2\sqrt{\jjrrCRN\jjRevrrCRN}$,
and a vector of (half)forces $\netforce\in\mathbb{R}^{M/2}$ as $\netforcerr=\frac{1}{2}\ln(\jjrrCRN/\jjRevrrCRN)$.
We use the notation $\barBBdiv$ to indicate the $\numstate\times\numedge/2$
matrix that only has columns for the forward  reaction ($\rho$) in each
pair $(\rho,\negedge)$. $\barBBdiv$ maps currents to time evolution
vectors: $\dtp=\barBBdiv\netflux=\BBdiv\jj$. We emphasize that Ref.~\citep{Kobayashi2022}
uses the convention that forces $\netforcerr=\frac{1}{2}\ln(\jjrrCRN/\jjRevrrCRN)$
are scaled by $1/2$ relative to the forces as defined in this paper,
$f_{r}=\ln(\jjrrCRN/\jjRevrrCRN)$.

Observe that the currents can be expressed as 
\[
\netjR=\omega_{r}\sinh(\netforcerr)=\sqrt{\jjrrCRN\jjRevrrCRN}\left(\sqrt{\frac{\jjrrCRN}{\jjRevrrCRN}}-\sqrt{\frac{\jjRevrrCRN}{\jjrrCRN}}\right)=\jjrrCRN-\jjRevrrCRN.
\]
Conversely, this equation can be solved for $\netforcerr$ as 
\begin{align*}
\netforcerr=\sinh^{-1}(\netjR/\omega_{r}).
\end{align*}
These relations can also be derived from a higher level structure.
Define two dual convex functions which are the Legendre conjugate
of each other: for a fixed $\bm{\omega}$, the convex function 
\[
\Psi_{\omega}(\netflux'):=\sum_{r}\left[\netjR'\sinh^{-1}(\netjR'/\omega_{r})-\omega_{r}\left[\sqrt{1+(\netjR'/\omega_{r})^{2}}-1\right]\right]
\]
is the Legendre conjugate of 
\begin{align*}
\Psi_{\omega}^{*}(\netforce')=\sum_{r}\omega_{r}\Big[\cosh(\netforcerr')-1\Big],
\end{align*}
and they specify the current and force across reaction $r$ as 
\begin{align*}
\netjR=\partial_{\netforcerr}\Psi_{\omega}^{*}(\netforce),\quad\netforcerr=\partial_{\netjR}\Psi_{\omega}(\netflux).
\end{align*}
Note that for any $\bm{\omega}$, $\Psi_{\omega}(\zz)=\Psi_{\omega}^{*}(\zz)=0$
holds and it is their minima. In general, a convex function $\varphi(\bm{x})$
leads to the Bregman divergence $D(\bm{x}\Vert\bm{x}'):=\varphi(\bm{x})-\varphi(\bm{x}')-\langle\bm{x}-\bm{x}',\nabla\varphi(\bm{x}')\rangle\geq0$,
where $\langle\cdot,\cdot\rangle$ is the normal inner product and
$\nabla\varphi(\bm{x})$ is the gradient vector $(\partial_{x_{1}}\varphi(\bm{x}),\partial_{x_{2}}\varphi(\bm{x}),\dots)^{T}$
of the function~\citep{amari2016information}. For a fixed $\bm{\omega}$,
we can define the Bregman divergences $D_{\omega}$ and the dual one
$D_{\omega}^{*}$ by 
\begin{align*}
D_{\omega}(\netflux'\Vert\netflux'') & \!:=\!\Psi_{\omega}(\netflux')\!-\!\Psi_{\omega}(\netflux'')\!-\!\langle\netflux'\!-\!\netflux'',\nabla\Psi_{\omega}(\netflux'')\rangle\\
D_{\omega}^{*}(\netforce'\Vert\netforce'') & \!:=\!\Psi_{\omega}^{*}(\netforce')\!-\!\Psi_{\omega}^{*}(\netforce'')\!-\!\langle\netforce'\!-\!\netforce'',\nabla\Psi_{\omega}^{*}(\netforce'')\rangle
\end{align*}
As a general property of Bregman divergences and the Legendre transformation,
we have 
\begin{align}
D_{\omega}(\netflux'\Vert\netflux'')=D_{\omega}^{*}(\netforce''\Vert\netforce')
\end{align}
when $(\netflux',\netforce')$ and $(\netflux'',\netforce'')$ are
Legendre dual coordinates. In this situation, we also have 
\begin{align}
D_{\omega}(\netflux'\Vert\netflux'')=\Psi_{\omega}(\netflux')+\Psi_{\omega}^{*}(\netforce'')-\langle\netflux',\netforce''\rangle,
\end{align}
which leads to 
\begin{align}
\epr=\langle\netflux,\netforce\rangle=\Psi_{\omega}(\netflux)+\Psi_{\omega}^{*}(\netforce).
\end{align}
Note that in general, these Bregman divergences cannot be expressed
as KL divergence because the current $\netjR$ can be negative. Therefore,
they cannot be related to the EPR, in the same way that we relate
nonnegative one-way fluxes to EPR via the divergence $\mathcal{D}$
in \mainref{eq:eprdist}.

The Hessian decomposition~\citep{Kobayashi2022} is defined by using
two special points: $(\netflux_{\eq},\netforce_{\eq})$, which represent
conservative currents/forces, and $(\netflux_{\sss},\netforce_{\sss})$,
which represent steady-state currents/forces. Given these two pairs
of currents/forces, we have 
\begin{align}
\eprHK^{\text{hess}}:=\Psi_{\omega}(\netflux_{\eq})+D_{\omega}^{*}(\netforce\Vert\netforce_{\sss}),\\
\eprEX^{\text{hess}}:=\Psi_{\omega}^{*}(\ff_{\sss})+D_{\omega}(\netflux\Vert\netflux_{\eq}).
\end{align}
To explain how $(\netflux_{\eq},\netforce_{\eq})$ and $(\netflux_{\sss},\netforce_{\sss})$
are determined, we define two kinds of sets. We define $\mathcal{P}(\netflux')$
as the set of currents that induce the same dynamics as $\netflux'$
by 
\begin{align}
\mathcal{P}(\netflux'):=\{\netflux''\in\mathbb{R}^{M/2}\mid\barBBdiv\netflux''=\barBBdiv\netflux'\}.
\end{align}
The other space $\mathcal{M}_{\omega}(\netforce')$ is defined as
the set of currents that are given by $\netforce'$ plus some conservative
forces: 
\begin{align}
\mathcal{M}_{\omega}(\netforce'):=\{\nabla\Psi_{\omega}^{*}(\netforce'')\mid\netforce''\in\netforce'+\mathrm{im}\,\barBBgrad\},
\end{align}
where $\netforce'+\mathrm{im}\,\barBBgrad:=\{\netforce'+\barBBgrad\pot\mid\pot\in\mathbb{R}^{N}\}$.
Then, $\netflux_{\eq}$ and $\netflux_{\sss}$ are given as unique
intersections as 
\begin{align}
\netflux_{\eq}:=\mathcal{P}(\netflux)\cap\mathcal{M}_{\omega}(\zz),\;\;\netflux_{\sss}:=\mathcal{P}(\zz)\cap\mathcal{M}_{\omega}(\netforce),
\end{align}
while the corresponding forces $\netforce_{\eq}$ and $\netforce_{\sss}$
are provided as $\nabla\Psi_{\omega}(\netflux_{\eq})$ and $\nabla\Psi_{\omega}(\netflux_{\sss})$.
Therefore, we see that, with frenetic activity being fixed, $\netflux_{\eq}$
is the current induced by a conservative force which recovers the
original dynamics, while $\netflux_{\sss}$ is the steady-state current
which is given by a force that has the same nonconservative contribution
as the actual force.

We note variational characterizations of $(\netflux_{\eq},\netforce_{\eq})$
and $(\netflux_{\sss},\netforce_{\sss})$, which can make easier to
calculate the decomposition numerically. $\netflux_{\eq}$ is given
by 
\begin{align}
\netflux_{\eq}=\argmin_{\netflux'\in\mathcal{P}(\netflux)}\Psi_{\omega}(\netflux'),
\end{align}
while $\netforce_{\sss}$ is obtained as 
\begin{align}
\netforce_{\sss}=\argmin_{\netforce'\in\netforce+\mathrm{im}\,\barBBgrad}\Psi_{\omega}^{*}(\netforce').
\end{align}

Next, let us focus on a specific chemical reaction network. In Ref.~\citep{Kobayashi2022},
the authors discuss the reaction network 
\begin{align}
2X\underset{k_{1}^{\leftarrow}}{\overset{k_{1}^{\rightarrow}}{\rightleftarrows}}2Y\underset{k_{2}^{\leftarrow}}{\overset{k_{2}^{\rightarrow}}{\rightleftarrows}}X+Y\underset{k_{3}^{\leftarrow}}{\overset{k_{3}^{\rightarrow}}{\rightleftarrows}}2X,\label{eq:appcrn}
\end{align}
assuming the mass action kinetics with rate constants presented in
the chemical equations. We calculate our EPRs $\eprHK$, $\eprEX$,
the Onsager EPRs $\eprHKons$, $\eprEXons$, and the Hessian EPRs
$\eprHK^{\text{hess}}$, $\eprEX^{\text{hess}}$, with the same parameters
as Ref.~\citep{Kobayashi2022}. Concretely, we used the rate constants
$k_{1}^{\rightarrow}=1/2,k_{1}^{\leftarrow}=2,k_{2}^{\rightarrow}=4,k_{2}^{\leftarrow}=47/4,k_{3}^{\rightarrow}=\sqrt{2},$
and $k_{3}^{\leftarrow}=15/2+2\sqrt{2}$ to obtain (a) in \cref{fig:compareeprs},
or $k_{1}^{\rightarrow}=1/2,k_{1}^{\leftarrow}=2,k_{2}^{\rightarrow}=1/17,k_{2}^{\leftarrow}=85/8,k_{3}^{\rightarrow}=273/68,$
and $k_{3}^{\leftarrow}=137/68$ to obtain (b). The three decompositions
are exhibited in \cref{fig:compareeprs}, which reproduces numerical
results obtained in Ref.~\citep{Kobayashi2022}. The inequality $\eprEX\leq\eprEXons$
is also verified. In addition, we observe numerically that $\eprEX\leq\eprEX^{\text{hess}}\leq\eprEXons$,
although we have not proved analytically that these inequalities hold
in general.

\vspace{15pt}

\begin{figure}[H]
\centering \includegraphics[width=3.5in]{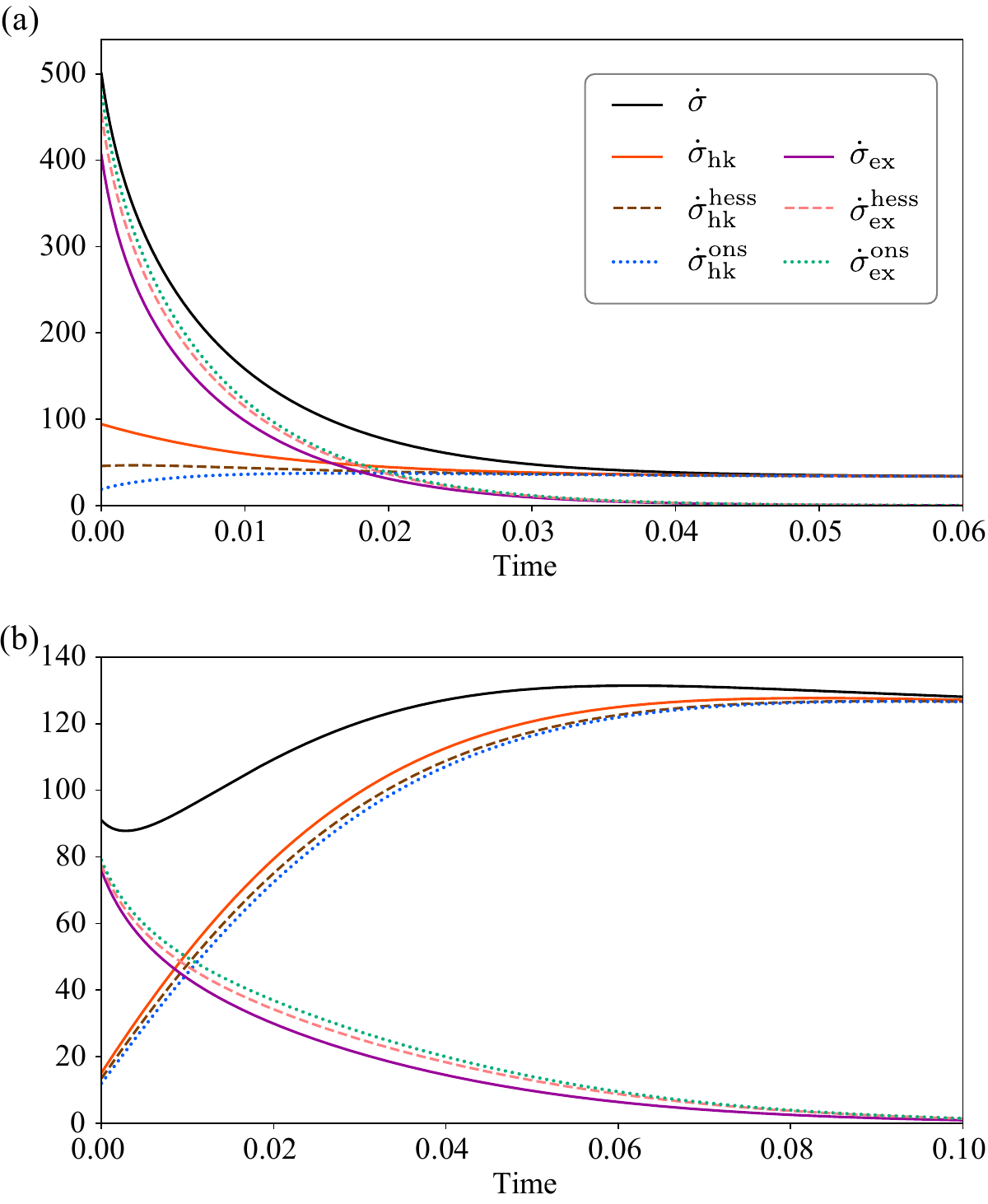} \caption{Comparison of three EPR decompositions. We calculate EPRs of the chemical
reaction network in \cref{eq:appcrn} for two distinct rate constants
(detailed values are given in the text).}
\label{fig:compareeprs} 
\end{figure}

\clearpage{}

\fi 

\end{document}